\documentclass[aps,prx,twocolumn,final,letterpaper]{revtex4}

\usepackage{appendix}
\usepackage{graphicx}   
\usepackage{import}                         
\usepackage{epstopdf}
\usepackage{amsmath} 
\usepackage{bm}
\usepackage{amssymb}
\usepackage{quotes}
\usepackage{transparent}
\usepackage{dcolumn}
\usepackage{multirow}
\usepackage{cancel} 
\usepackage{mdframed}
\usepackage{color}
\usepackage{bm}
\usepackage{dsfont}
\usepackage{slashed}
\usepackage{braket}

\newcommand{\rv}{\mathbf{r}}

\newcommand{\kv}{\mathbf{k}}

\newcommand{\appropto}{\mathrel{\vcenter{
  \offinterlineskip\halign{\hfil$##$\cr
    \propto\cr\noalign{\kern.2pt}\sim\cr\noalign{\kern-2.5pt}}}}}

\renewcommand{\Re}{\operatorname{Re}}
\renewcommand{\Im}{\operatorname{Im}}

\newcommand{\Jv}{\mathbf{J}}

\newcommand{\Ev}{\mathbf{E}}
\newcommand{\Dv}{\mathbf{D}}

\newcommand{\dv}{\mathbf{d}}

\newcommand{\curl}{\nabla\times}

\begin{document}
\rmfamily

\title{Optical properties of dispersive time-dependent materials}
\author{Jamison Sloan$^{1,2}$}
\email{jamison@mit.edu}
\author{Nicholas Rivera$^{3,4}$}
\author{John D. Joannopoulos$^{3}$}
\author{Marin Solja\v{c}i\'{c}$^{3}$}

\affiliation{$^1$ Department of Electrical Engineering and Computer Science, MIT, Cambridge, MA 02139, USA\looseness=-1}
\affiliation{$^2$ Research Laboratory of Electronics, MIT, Cambridge, MA 02139, USA\looseness=-1}
\affiliation{$^3$ Department of Physics, MIT, Cambridge, MA 02139, USA\looseness=-1}
\affiliation{$^4$ Department of Physics, Harvard University, Cambridge, MA 02138, USA\looseness=-1}

\noindent	

\begin{abstract}
Time-varying optical materials have attracted recent interest for their potential to enable frequency conversion, nonreciprocal physics, photonic time-crystals, and more. However, the description of time-varying materials has been primarily limited to regimes where material resonances (i.e., dispersion) can be neglected. In this work, we describe how the optics of these dispersive time-varying materials emerges from microscopic quantum mechanical models of time-driven systems. Our results are based on a framework for describing the optics of dispersive time-varying materials through quantum mechanical linear response theory. Importantly, we clarify how response functions for time-varying materials are connected to energy transfer. We provide three examples of our framework applied to systems which can be used to model a wide variety of experiments: few level models that can describe atoms, spins, or superconducting qubits, oscillator models which can describe the strong response of polar insulators, and strongly driven atom models which can describe the highly nonperturbative optical response of materials undergoing high harmonic generation. We anticipate that our results will be broadly applicable to electromagnetic phenomena in strongly time-varying systems.
\end{abstract}

\maketitle

\section{Introduction}




The propagation of electromagnetic waves through materials represents an essential component of light-matter interactions, and lies at the heart of countless physical phenomena and technological applications. In many bulk materials, the dominant features of electromagnetic wave propagation can be described by a simple  complex refractive index which encodes the speed of wave propagation, as well as the rate of dissipation. In fact, the ability to describe a complex many-body system such as a solid with a frequency dependent refractive index is critical for a practical description of many systems. Over the last century, a great deal of effort has gone into understanding the origins and fundamental properties of optical response, leading to important devices such as detectors, LEDs, solar cells, and lasers. Nowadays, artificial structures such as photonic crystals, layered 2D materials, and metamaterials are routinely created to provide further control over optical response, leading to increased command over the interactions between light and matter.

Many of the basic assumptions about the nature of optical response and wave propagation rely on considering optical materials as time translation-invariant --- the same at all times. However, a recent surge of interest has developed in the possibilities that may be enabled by materials which break this assumption --- in other words, materials which vary in time \cite{galiffi2021photonics}. In practice, time-varying materials are typically created by applying strong temporal modulations to stationary materials in the form of external fields. These time-varying materials may exhibit rich physics such as frequency conversion \cite{zhou2020broadband}, scattering from temporal interfaces \cite{pacheco2020temporal, plansinis2015temporal}, nonreciprocity \cite{shaltout2015time, li2022nonreciprocity}, and amplification \cite{pendry2020new}. Moreover, a recent interest has sparked in the study of so-called ``photonic time crystals'' which have a strong temporally periodic index variation, enabling new directions in topological physics \cite{lustig2018topological} and light-matter interactions \cite{dikopoltsev2022light, lyubarov2022amplified}. In the context of metamaterials, time has recently been identified as additional degree of freedom which can be added to create ``spatiotemporal metamaterials'' \cite{caloz2019spacetime, engheta2021metamaterials}. Additionally, the possibility of strongly time-modulated materials introduces new fundamental questions about the nature of quantum light-matter interactions in time-modulated systems, including control over the generation of entangled photon pairs \cite{yablonovitch1989accelerating, sloan2020casimir, kort2021space}.


To achieve these goals, it will become increasingly important to accurately describe the optical response of time-varying materials in the most general setting. Past work on time-varying media has typically assumed that the time modulations to a material occur at frequencies away from intrinsic resonances in the material \cite{law1994effective, lustig2018topological, zurita2009reflection,chu1972wave,harfoush1991scattering,fante1971transmission,holberg1966parametric}. In these cases, it is sufficient to consider a permittivity $\varepsilon(t)$ which is nondispersive, associated with an instantaneous polarization response. However, there are many systems, especially those which are varied quickly in time, which are not adequately accounted for by this framework. For example, the description of wave propagation on a strongly driven conductor requires the simultaneous description of driving and plasmonic dispersion. In fact, understanding the influence of dispersion has recently been identified as a key challenge in the field of time-varying materials \cite{solis2021functional}. Some semiclassical models have been proposed for particular systems \cite{ptitcyn2021scattering, solis2021functional}. Yet, there is still no comprehensive framework for describing the optical physics of dispersive time dependent materials from first principles.

It is tempting to take a theoretical model (or data) for the optical response of an undriven system, and then introduce time dependence. While this is a valid approach for slow variations, it is not reliable in general. The key issue is that in a strongly time-modulated system, the optical response depends on the new microscopic dynamics of the driven system, which will in general not be adequately captured by this approach. Therefore, the optical response of time-driven materials should ideally be considered on the basis of first principles, starting from a microscopic description of the driven system.

In this work, we present a general framework which describes the optics of dispersive time-dependent materials based on microscopic quantum mechanical dynamics. By doing so, we answer a fundamental question about the nature of energy transfer in time-varying systems, namely the significance (or in some cases, lack thereof) of the imaginary part of response functions. We also specialize many of our results to the particularly intriguing case of time-periodic (i.e. ``Floquet'') systems. In this case, many of our results are simplified by the use of Floquet theory to describe both the quantum and macroscopic electromagnetic aspects of problems. We provide examples of our framework across variety of systems: time-modulated superconducting qubits in the GHz, time-modulated polar insulators with optical phonon resonances, and strongly driven gasses which exhibit high harmonic generation (HHG). Our framework, when applied to these systems, enables us to discover a wide range of physics such as pulse propagation in dispersive photonic time-crystals, nonperturbative frequency conversion, and energy loss/gain. 

These findings point toward a future where time-varying linear response theory is an important theoretical and experimental tool for studying time-varying optical materials. In an experimental setting, a great benefit of using linear response theory in these complex systems is that linear response functions can be measured, rather than computed. Another advantage of this framework is that it allows one to characterize nonperturbative nonlinearities, which can be important in systems which are very strongly driven.

The organization of this work is as follows: In Section \ref{sec:theory}, we give an overview of our framework which incorporates quantum mechanical linear response theory, and classical optics to describe wave propagation in dispersive time-varying materials. This section includes an important discussion about Kramers-Kronig relations, and how response functions encode energy transfer in time-varying optical systems. We also summarize some of the important simplifications when these results are specialized to time-periodic (Floquet) systems, leading to a compact description of wave propagation in dispersive photonic time-crystals. After describing the key foundations, we provide three distinct examples of our framework applied to different types of microscopic models for driven systems. In Section \ref{subsec:2LS}, we describe wave propagation and energy transfer in a material whose optical response is characterized by time-modulated two-level systems. Intriguingly, we find that in the presence of sufficiently strong modulation, this type of system can exhibit resonant gain in its ground state. Such a model is relevant for describing metamaterials which could be formed from networks of superconducting qubits. In Section \ref{subsec:LPO}, we describe the optical response and resultant scattering processes in a system described by a time-varying Lorentz oscillator model which we refer to as a ``Lorentz parametric oscillator.'' Such a model is relevant for describing time-modulated polar insulators with an infrared optical response which is dominated by optical phonon resonances. Finally, in Section \ref{subsec:HHG}, we describe the highly nonperturbative frequency conversion which may occur in gases undergoing high harmonic generation (HHG). This result paves the way toward using strongly time-driven materials to create artificial optical response at ultraviolet and X-ray frequencies.



\begin{figure*}
    \centering
    \includegraphics{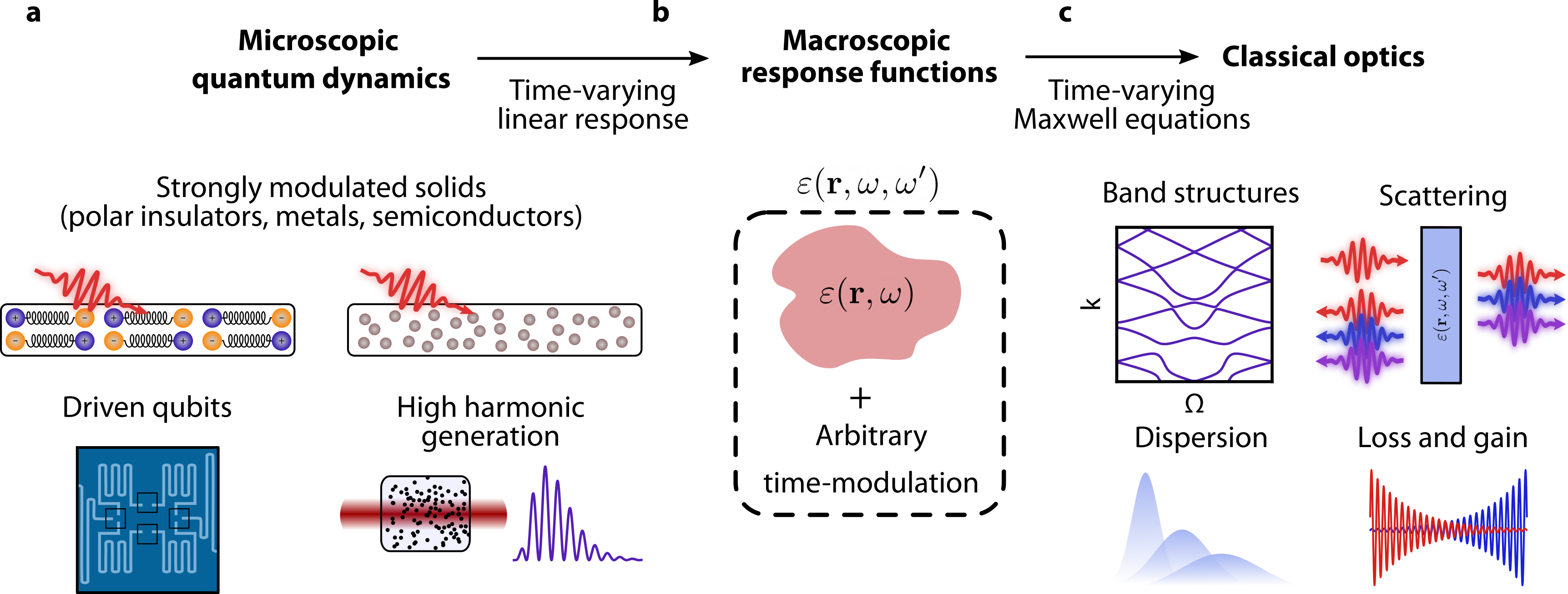}
    \caption{\textbf{General framework for describing the optics of dispersive and strongly time-dependent systems.} (a) Examples of time-dependent quantum mechanical systems whose optical response requires time-varying linear response theory. (b) Models of these microscopic dynamics can then be used to construct macroscopic response functions which may also vary spatially to account for material structures. For example, a dispersive dielectric structure $\varepsilon(\rv, \omega)$ in the absence of time-modulation can be described in terms of a two-frequency response function $\varepsilon(\rv, \omega, \omega')$ in the presence of time-modulation. (c) These response functions can be incorporated into the Maxwell equations to describe optical features of these systems, such as ``free'' wave propagation, scattering, and energy transfer.}
    \label{fig:schematic}
\end{figure*}

\section{Theoretical Framework}
\label{sec:theory}

In this section, we describe our general framework for constructing new time-dependent optical materials from microscopic quantum mechanical models (Fig. 1). In such models, the dynamics are described by the solution to the Schr\"{o}dinger equation with a time-dependent Hamiltonian $H(t)$. Generally, these microscopic dynamics can depend on many-body effects in a complicated manner. In this work, we will focus on materials which are well-described by constructing an effective bulk response from a collection of single particle dynamics; however, many of our conclusions hold more broadly. Once the relevant Schrodinger equation has been solved, the dipole response functions of single particles can be constructed, and then transformed into bulk macroscopic response functions such as $\varepsilon(\omega, \omega')$. In systems where dissipation mechanisms are important, this process can also be followed by solving an appropriate master equation which rigorously incorporates the dissipative dynamics.

With a macroscopic response function in hand, one can then use classical electrodynamics to describe wave propagation and energy transfer in dispersive time-varying media. For example, strongly modulated systems which are periodic in time (photonic ``time crystals'') can be associated with a band structure which indicates the relationship between wavevector and quasi-frequency in the driven material (see dispersion diagram in Fig. \ref{fig:schematic}c). In systems with strong light-matter hybridization, this provides a direct way to solve for the polaritons of the driven system. Another example is the use of time-dependent response functions to compute frequency-dependent scattering from a structure such as a thin film of a time-dependent material. Experiments of this type have been performed on epsilon-near-zero (ENZ) materials \cite{alu2007epsilon, reshef2019nonlinear} which have demonstrated so-called temporal refraction \cite{zhou2020broadband}. The linear response formulation that we detail in this work allows for the prediction of these behaviors in systems where dispersion is critical.


\subsection{Time-varying linear response theory}

The foundation of our approach is time-varying linear response theory, which describes how some observable of a time-varying quantum mechanical system evolves due to the presence of a weak probe field \cite{kubo1957statistical}. For the context of this discussion, we will focus on the polarizability which dictates how an electric field probe $\Ev(t)$ induces a change to the dipole moment $\Braket{\delta\mathbf{d}(t)}$, where $\Braket{\cdot}$ denotes a quantum mechanical expectation value. While we will focus on the polarizability of a single point-like particle, these concepts apply equally well to other response functions such as susceptibility, permittivity, conductivity, magnetic permeability, etc. 

A dispersive time-driven material must in general be described with response functions which refer to two times (or two frequencies) \cite{landau2013electrodynamics}. This is due to the fact that the system is not time-translation invariant, and thus memory effects depend on the absolute time at which a probe interacts with the system of interest. For the polarizability example, the change to the dipole moment can be written in terms of the two-time polarizability $\boldsymbol\alpha(t, t')$ tensor and the probe electric field $\Ev(t)$ as 
\begin{subequations}
\begin{align}
    \Braket{\delta\dv(t)} &= \int dt'\, \boldsymbol\alpha(t, t') \Ev(t') \\
    \Braket{\delta\dv(\omega)} &= \int \frac{d\omega'}{2\pi} \, \boldsymbol\alpha(\omega, \omega') \Ev(\omega'),
\end{align}
\end{subequations}
where frequency domain polarizability is defined as:
\begin{equation}
    \boldsymbol\alpha(\omega, \omega') = \int dt\,dt'\, e^{i\omega t} e^{-i\omega't'} \boldsymbol\alpha(t,t').
\end{equation}
The phase convention on the Fourier transform is chosen so that in the time-independent limit, $\boldsymbol\alpha(\omega, \omega') = 2\pi\boldsymbol\alpha(\omega)\delta(\omega-\omega')$; hence, the usual relation $\Braket{\delta\dv(\omega)} = \boldsymbol\alpha(\omega)\Ev(\omega)$ is recovered. 

The response function of any time-dependent system must be characterized by its temporal microscopic dynamics. In the case of a time-dependent point-like particle, the polarizability is given by the Kubo formula
\begin{equation}
    \boldsymbol\alpha(t, t') = \frac{i}{\hbar}\theta(t-t')\Braket{[\dv(t), \dv(t')]},
    \label{eq:time_domain_kubo}
\end{equation}
where $\dv(t)$ is the interaction picture dipole operator, and the expectation value is taken in the initial state of the system. In solid state systems where many-body effects are important, the time-varying dielectric function or conductivity can be appropriately formulated in a similar way \cite{rudner2020floquet, wackerl2020floquet}.



\emph{Causality and K.K. relations:} In time-independent systems, the requirement of passivity is tightly linked to the possible forms of a generic response function $\chi(\omega)$ via the Kramers-Kronig (K.K.) relations \cite{kramers1927diffusion, kronig1926theory}. When time-dependence is introduced, this passivity requirement dissolves, as the drive provides energy to the system that can lead to gain, among other effects. Despite this added complexity, time-varying response functions are still constrained by causality. Specifically, any time-dependent linear response function $\chi(t, t')$ must obey the relationship $\chi(t, t') = \theta(t - t')\chi(t, t')$ so that changes to an observable at time $t$ are only caused by interactions at times $t' < t$. In the frequency domain, we show (Appendix \ref{subsec:kk_appendix}) this requirement leads to the K.K. relationship:
\begin{equation}
    \chi(\omega, \omega') = \frac{i}{\pi}\mathcal{P}\int d\omega'' \frac{\chi(\omega - \omega'', \omega' - \omega'')}{\omega''},
    \label{eq:time_dep_kk}
\end{equation}
where $\mathcal{P}$ denotes the principal value. By taking real and imaginary parts of Eq. \ref{eq:time_dep_kk}, one can obtain a direct relationship between the real and imaginary parts of $\chi(\omega, \omega')$. In the limit that the material is time translation-invariant, the response function takes the limiting form $\chi(\omega, \omega') \to \chi(\omega) \cdot 2\pi\delta(\omega - \omega')$, and the standard K.K. relation is recovered. 

\emph{Energy transfer:} In a time-independent material, the energy absorbed by the material from a passing wave at frequency $\omega$ is proportional to $\Im \chi(\omega)$. These time-dependent K.K. relations raise an immediate question about energy transfer in time-dependent materials: does $\Im \chi(\omega, \omega')$ still encode energy dissipation (or gain) for a time-dependent material? We will show here that this is generally not the case.

To do so, we consider the work done by a probe field on a time-dependent polarizable particle. The total energy transferred to a point dipole can be written as $U = \int_0^\infty d\omega P(\omega)$. In this expression, $P(\omega) = -\frac{\omega}{\pi}\Im \left[\mathbf{d}(\omega) \cdot \mathbf{E}^*(\omega)\right]$ is the energy per unit frequency dissipated, $\mathbf{d}(\omega)$ is the dipole moment, and $\mathbf{E}(\omega)$ is the probe field. Assuming the point dipole is described by the polarizability $\boldsymbol\alpha(\omega, \omega')$, the energy dissipated per frequency is:
\begin{equation}
    P(\omega) = -\frac{\omega}{\pi} \Im \int_{-\infty}^\infty \frac{d\omega'}{2\pi}\mathbf{E}^*(\omega)\boldsymbol\alpha(\omega, \omega')\mathbf{E}(\omega').
    \label{eq:power_absorbed}
\end{equation}
Unlike the equivalent expression for time-independent media, Eq. \ref{eq:power_absorbed} does not posses a clear sign; this is consistent with the general feature of time-dependent media that a passing wave can lose or gain energy \cite{pendry2021gain, galiffi2022archimedes} (and in fact, we will show cases where $P > 0$, in violation of passivity). Moreover, the amount of energy lost or gained can in general depend on the phase of passing waves. To see this explicitly, consider that for a monochromatic probe $E(t) = E_0 \cos(\omega_p t - \phi)$ of a single polarization incident upon an isotropic particle, the total energy dissipated is expressed as 
\begin{equation}
    U = \frac{E_0^2 \omega_p}{2}\Im\left[\alpha(\omega_p, \omega_p) + e^{-2i\phi}\alpha(\omega_p, -\omega_p)\right].
    \label{eq:monochromatic_probe}
\end{equation}
The form of Eq. \ref{eq:monochromatic_probe} explicitly shows a contribution to the energy loss/gain which depends on the phase $\phi$ of the probe. This is, for example, exactly the type of physics exemplified by an optical parametric oscillator, where the signal can either either be exponentially amplified or attenuated depending on the phase of the input. Later, we give examples of systems where $U$ can take on either sign, depending on whether the energy of the probe is absorbed or amplified.

For a monochromatic probe, there are two contributions to the change in energy, corresponding to the frequency components of the probe at $\pm \omega_p$. The first contribution comes from $\Im \alpha(\omega_p, \omega_p)$, which is independent of the probe phase. For this term, the positive frequency components of the probe field induce a dipole moment at that same positive frequency. It is this term which reduces to the usual relation that energy transfer is proportional to $\Im \alpha(\omega)$ in a time-independent system. The second contribution comes from $\Im [e^{-2i\phi} \alpha(\omega_p, -\omega_p)]$, which depends explicitly on the temporal relationship between the probe field and dynamics of the driven system through the phase $\phi$. For this term, the negative frequency component of the probe $-\omega$ is shifted up by $2\omega_p$ to $\omega_p$. From this, we can see that dissipated energy can depend on both the real and imaginary parts of the response function $\chi(\omega, \omega')$.



\subsubsection{From microscopic to macroscopic}

The time-varying response functions discussed here are useful not only for describing energy transfer into a medium, but also wave propagation. To see this, we consider the construction of potentially spatially and temporally varying response functions which are used to describe some time driven photonic structure. For many systems, the point-like particles described by a time-dependent polarizability $\alpha(\omega, \omega')$ can be used to describe the optical response of bulk materials, as is routinely considered for time-independent media. In the simplest possible case, one assumes that the polarizable particles are packed with a volume density $n$, allowing one to define a unitless susceptibility $\chi(\omega, \omega')$ by $\chi(\omega, \omega') = (n/\varepsilon_0) \alpha(\omega, \omega')$. Such a scheme neglects local field effects, which can be accounted for using a Clausius-Mossotti relation, or similar method which is appropriate to the geometry \cite{aspnes1982local}. Spatial arguments can also be incorporated in the case that the material structure varies spatially, or if the time-varying material possesses some joint spatio-temporal evolution. Such modulations have been recently considered in the context of constructing ``spatiotemporal metamaterials'' \cite{caloz2019spacetime} and ``spatiotemporal photonic crystals'' \cite{sharabi2022spatiotemporal}.

We can thus write a form of Maxwell's equations in frequency space which uses a two-frequency linear response function in its constitutive relation. To do so, it is helpful to define a permittivity $\varepsilon(\rv, \omega, \omega') = 2\pi\delta(\omega-\omega') + \chi(\rv, \omega, \omega')$ which relates the displacement and electric fields as $\Dv(\rv, \omega) = \varepsilon_0 \int \frac{d\omega}{2\pi}\varepsilon(\rv, \omega, \omega')\Ev(\rv, \omega')$. Under this definition, the electric field $\Ev(\rv, \omega)$ in the presence of a current source $\Jv(\rv, \omega)$ obeys:
\begin{equation}
\begin{split}
    \curl\curl\Ev(\rv, \omega) - \frac{\omega^2}{c^2}\int\frac{d\omega'}{2\pi}&\varepsilon(\rv, \omega, \omega') \Ev(\rv, \omega') \\
    &= i\omega\mu_0\Jv(\rv, \omega).
\end{split}
\end{equation}
Due to the time-dependence, this form of the Maxwell equation is nonlocal in frequency space. In general, FDTD methods may be required in order to simulate full spatial and temporal dependencies of time-driven systems. However, we show in the next section that for time-periodic systems, the linear response functions take a form which enable significant simplifications.

\subsection{Specialization to time-periodic systems}


One general class of time-modulated systems which is of particular interest are those in which the modulation is periodic in time \cite{fante1971transmission, morgenthaler1958velocity}. In certain cases, such systems have been termed ``photonic time crystals'' (PTCs). Most PTCs considered so far have been described in terms of a ``nondispersive'' permittivity $\varepsilon(t) = \varepsilon(t + T)$, where $T$ is the period. Furthermore, for materials to be considered PTCs, it is usually assumed that the relative temporal variations to the material are substantial ($\Delta\varepsilon \gtrsim 0.1$) so that the nature of wave propagation departs substantially from that in an unmodulated counterpart. Due to their periodic nature, a spatially homogeneous PTC can be associated with a band structure which relates wavenumbers $k$ to \emph{quasifrequencies} $\Omega$, which lie in a temporal Brillouin zone (TBZ) set by $\Omega_0 \equiv 2\pi/T$. This phenomenon is well studied, and manifests many natural analogies to spatial photonic crystals. 

As an important extension of these ideas, we introduce the concept of dispersive PTCs which result from temporal modulations which are fundamentally dispersive. In this section, we specialize key results from above to time-periodic systems. The key result is that in a time-periodic system, a response function $\chi(\omega, \omega')$ can be reduced to an integer series of response functions $\chi_k(\omega)$. 

\emph{Form of response functions:} In time periodic systems, the harmonic nature of the problem places constraints on the form of the response functions. Specifically, periodicity imposes the time-domain constraint $\chi(t, t') = \chi(t + T, t' + T)$. This immediately dictates that the frequency response is described by an integer series of response functions $\chi_k(\omega)$, which are defined such that
\begin{equation}
    \chi(\omega, \omega') = \sum_{k=-\infty}^\infty \chi_k(\omega) \cdot 2\pi\delta(\omega - \omega' + k\Omega_0).
    \label{eq:harmonic_response_function}
\end{equation}
From this form, we see that $\chi_k(\omega)$ encodes how an applied field at frequency $\omega$ induces a response at frequency $\omega + k\Omega_0$.

\emph{Kramers-Kronig relations, energy transfer:} The K.K. relations take a more familiar form in time-periodic media. By assuming that $\chi(\omega, \omega')$ obeys Eq. \ref{eq:harmonic_response_function}, we use Eq. \ref{eq:time_dep_kk} to deduce that the response function for each integer harmonic obeys the usual time-independent K.K. relation:
\begin{equation}
    \chi_k(\omega) = \frac{i}{\pi}\mathcal{P}\int d\omega'\frac{\chi_k(\omega')}{\omega - \omega'}.
    \label{eq:harmonic_kk}
\end{equation}
For time-periodic media, the integer order response functions allow for a particularly informative description of loss and gain. In particular, $\Im \alpha_0(\omega)$ encodes loss and gain which is independent of the probe phase, while $\alpha_k(\omega)$ of nonzero order encode loss and gain which depend on the probe phase. If we send a monochromatic probe field at such a material, the energy dissipated $U$ (from Eq. \ref{eq:monochromatic_probe}) reduces to 
\begin{equation}
\begin{split}
    \frac{U}{T} &= \frac{E_0^2 \omega_p}{2} \Im \alpha_0(\omega_p) \\
        &+ \frac{E_0^2 \omega_p}{2}\sum_{k=1}^\infty \Im\left[\alpha_k(\omega_p)e^{-2i\phi}\right]\delta_{2\omega_p = k\Omega_0} .    
\end{split}
    \label{eq:energy_transfer_periodic}
\end{equation}
This equation reveals several key pieces of information about the nature of energy transfer in time-periodic systems. We discuss these features by examining the two terms.

(1) In the first term, the imaginary part of the zeroth harmonic response function $\Im\alpha_0(\omega)$ \emph{carries unambiguous information about energy transfer which does not depend on the phase of the probe field.} Physically, $\alpha_0(\omega)$ encodes the polarization which is created at the same frequency as the probe, and is thus most closely connected to the dispersive response function of the undriven medium. Moreover, unlike in a ground state time-independent system, $\Im \alpha_0(\omega)$ is not restricted to be positive. This is analogous to an active medium which is pumped to an excited state, which can exhibit gain as characterized by a negative imaginary part of some response function. Instead here, the passivity can be broken by time-dependence rather than a population inversion. 

(2) In the second term, response function of orders $k \geq 1$ can affect the dissipated energy through phase-dependent effects under a special resonance condition. This resonance condition is indicated by a Kronecker delta function which requires that $2\omega_p = k\Omega_0$ for integer $k$. Physically, this condition results from negative frequency probe field components $-\omega_p$ which create polarization at frequencies which are shifted over by an integer number of harmonics: this positive frequency polarization can then interact with the positive frequency component $\omega_p$ of the probe field to do work (positive or negative). It is worth noting that this is the same type of mechanism responsible for parametric amplification processes as described in nonlinear optics, which are known to be sensitive to phase \cite{boyd2019nonlinear}.

For a given $\omega_p$ and $\Omega$, only one value of $k$, if any, can satisfy this condition. This potential additional term contains a phase dependence within the $\Im$ operator. This leads us to conclude that \emph{the real and imaginary parts of $\alpha_k(\omega)$ for $k \neq 0$ have no unique physical significance as far as energy transfer is concerned.} Thus the sign of $\Im\left[\alpha_k(\omega) e^{-2i\phi}\right]$, and the sign of the energy transfer, depends on the phase relationship between the probe field and the underlying microscopic dynamics that govern $\alpha_k(\omega)$. 



\emph{Eigenmodes in periodic systems:} In a time-periodic system, we can seek solutions to the sourceless Maxwell equation in a bulk medium. Due to the time periodicity, and spatial translation invariance, the Maxwell-Floquet modes take the form $\Ev(\rv, t) = e^{i \kv\cdot\rv}\sum_n u_{\Omega,n} e^{-i(\Omega+n\Omega_0)t}$, where $\kv$ is the wavevector, $\Omega$ is a quasifrequency in the first temporal Brillouin zone $[-\frac{\pi}{T}, \frac{\pi}{T})$, and $u_{\Omega, n}$ are a sequence of coefficients. In a medium with permittivity $\varepsilon(\omega, \omega) = 2\pi\delta(\omega - \omega') \varepsilon_{\text{bg}}(\omega) + \sum_k \Delta\chi_k(\omega) 2\pi\delta(\omega - \omega' + k\Omega_0)$, the amplitude of the wavevector $|\kv| = k_\Omega$ and coefficients can be found for a given quasifrequency by solving the eigenvalue problem:
\begin{equation}
	\frac{\Omega_n^2}{c^2}\varepsilon_{\text{bg}}(\Omega_n)u_n + \frac{\Omega_n^2}{c^2}\sum_m \Delta\chi_m(\Omega_n)u_{n-m} = k_\Omega^2 u_n,
	\label{eq:maxwell_floquet_eigenvalue}
\end{equation}
where $\Omega_n \equiv \Omega + n\Omega_0$ is the quasifrequency shifted by $n$ harmonics. This relation can be cast into a linear matrix problem which yields the band structures of dispersive photonic time crystals, examples of which are shown later in the text. We note that in the presence of very large loss or gain, an eigenmode expansion may not strictly form a complete basis for the set of possible solutions. However, in many cases, the eigenmode expansion may still provide accurate information about the dispersion relation. In systems where this approximation breaks down, Green's function methods can be employed to describe the propagation of waves from sources.

\section{Example Systems}

\subsection{Two-level system}
\label{subsec:2LS}

\begin{figure*}
    \centering
    \includegraphics{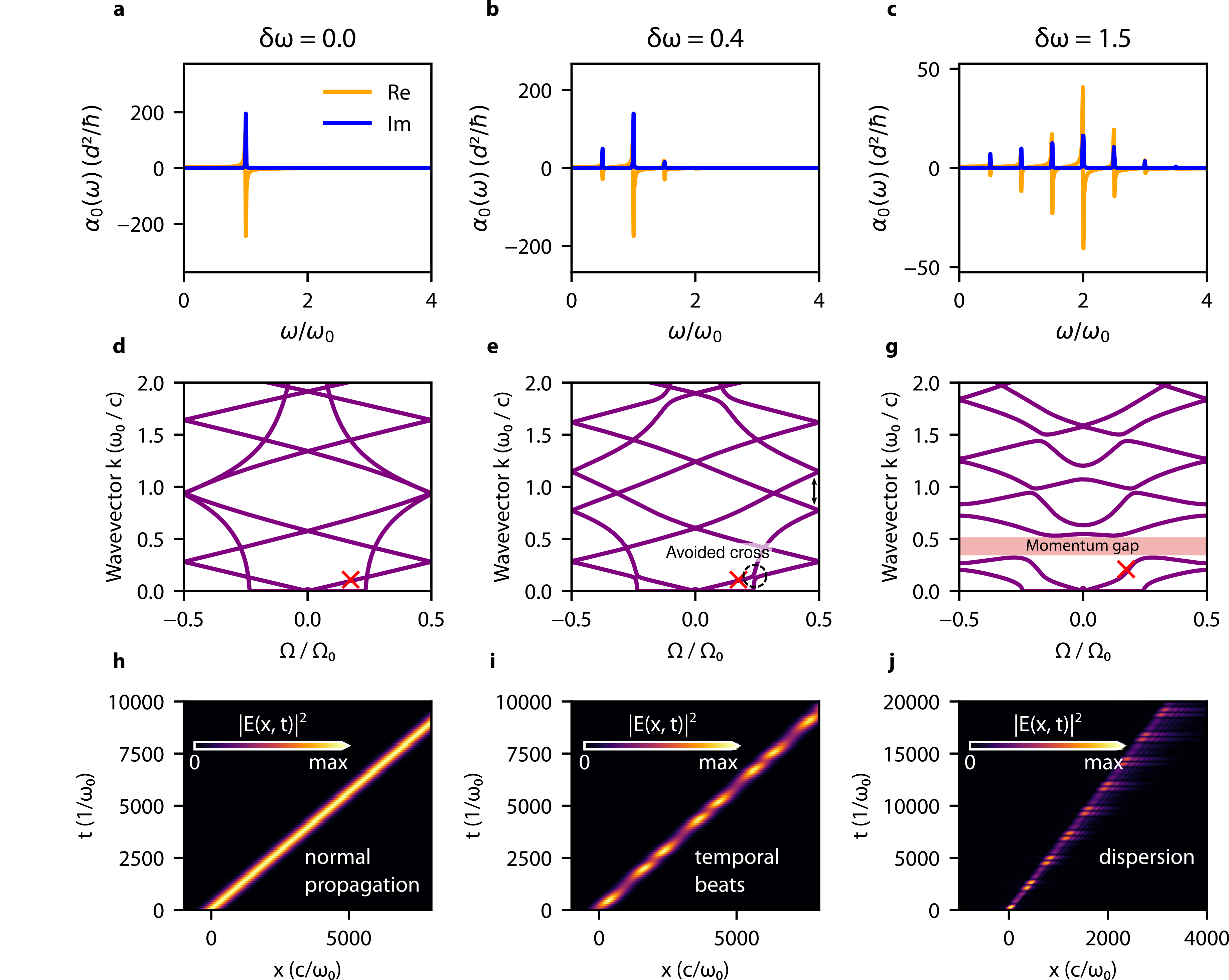}
    \caption{\textbf{Wave propagating in a time-driven two-level system from perturbative to nonperturbative regimes.} (a-c) Real and imaginary parts of zeroth order polarizability $\alpha_0(\omega)$ as a function of frequency for three different drive strengths $\delta\omega/\omega_0 = \{0, 0.4, 1.5\}$ and driving frequency $\Omega_0 = \omega_0 / 2$. Loss parameter is taken to be $\Gamma = 10^{-3} \omega_0$ for all peaks. (d-f) Dispersion relations for a time-periodic bulk medium which is composed of particles described by the polarizabilities $\alpha_k(\omega)$. Dispersion relations are plotted as a function of the quasifrequency $\Omega$ which lies in the temporal Brillouin zone $-\Omega_0/2 < \Omega \leq \Omega_0/2$. Sufficiently strong driving causes a gap to open in the momentum (panel g). (h-j) Propagation of a Gaussian wavepacket constructed from the modes indicated by a red $\times$ in corresponding panels (d-g). The line traced out in $(x,t)$ space indicates the group velocity of the packet. As the strength of the time modulation increases, new features such as temporal beating due to interference of harmonics, and wavepacket spreading due to group velocity dispersion.}
    \label{fig:2LS}
\end{figure*}

In this section, we discuss a two-level system (2LS) which has its energy splitting modulated strongly in time. When the modulation frequencies are close to the splitting frequency itself, the dispersive framework outlined above is required to describe linear response correctly. We focus specifically on a system with a Hamiltonian $H_{\text{2LS}}(t) = \frac{\omega_0}{2}\left(1 + \delta\omega\cos(\Omega_0 t)\right)\sigma_z$. Since the Hamiltonian is periodic in time, we can use the insight of Floquet theory that the system should behave like a stationary system, but with a ladder of quasi-energy levels. Physical systems which have realized Hamiltonians of this type include driven spins \cite{jiang2021floquet}, superconducting qubits \cite{deng2015observation}, quantum dots \cite{stehlik2016double}, and strongly modulated Rydberg atoms \cite{clark2019interacting}. A particularly appealing aspect of microwave schemes is that very strong modulations can be readily induced, leading to the nonperturbative regime of effects which do not fall under the purview of perturbative nonlinear optics. While we focus here on the $\sigma_z$-type modulation of a two-level system, many of the principles discussed here should carry over naturally to other types of two-level modulations, as well as systems with more discrete levels. 


While an undriven 2LS is restricted to make ground-excited transitions at the bare resonance frequency $\omega_0$, the time dependent 2LS we describe here can make transitions separated by harmonics of the drive. Thus, in a system which is well described in terms of a few energy levels, the driven system can exhibit optical response at frequencies different than that of the undriven system. Using the Floquet states of the Hamiltonian, and the Kubo formula, we find the single-particle polarizability:
\begin{equation}
    \alpha_k(\omega) = \frac{d^2}{\hbar}\Braket{\sigma_z}_0 \sum_n \left[\frac{J_{n+k} J_n}{\omega - \omega_n + i\Gamma_n} - \frac{J_{n-k} J_n}{\omega + \omega_n + i\Gamma_n}\right].
    \label{eq:2LS_response}
\end{equation}
Here, $\omega_n \equiv \omega_0 + n\Omega_0$ is the bare transition frequency shifted by $n$ harmonics, $J_n \equiv J_n(\delta\omega/\Omega_0)$ is the $n$-th order Bessel function evaluated at the driving strength parameter $\delta\omega/\Omega$, $\Gamma_n$ is the linewidth associated with each transition between Floquet levels, and $\Braket{\sigma_z}_0$ is an expectation value taken in the equilibrium state. More discussion about this equilibrium as well as the damping rates is shown in the Appendix. 

As the fractional change in the frequency $\delta\omega$ increases, the optical response of the system moves from a perturbative to nonperturbative regime (Figs. \ref{fig:2LS}a-c). The plots show the response function of zeroth harmonic order $\alpha_0(\omega)$ for three modulation strengths $\delta\omega/\omega_0 = 0, 0.4, 1.5$, and with a modulation frequency $\Omega_0 = \omega_0/2$. This response function encodes the amplitude with which a probe at frequency $\omega_p$ generates a change in the dipole moment at that same frequency. With no modulation, this describes the polarizability of a static 2LS, which is given by a Lorentzian (Fig. \ref{fig:2LS}a). For a stronger drive ($\delta\omega/\omega_0 = 0.4)$, sidepeaks emerge, which indicate optical response at $\omega_0 \pm \Omega_0$. At this strength of modulation, only the first harmonic contributes substantially, although others are present at levels which are not yet apparent in Fig. \ref{fig:2LS}b. This behavior shifts as the modulation strength nears or exceeds the static energy splitting $\omega_0$. Fig. \ref{fig:2LS}c shows an example of the nonperturbative regime, in which multiple harmonic orders are relevant. In this extreme limit, the strongest optical response actually occurs at $2\omega_0$, indicating that absorptions several steps up the Floquet ladder occur more strongly than the transition at $\omega_0$. We also note that the overall magnitude of the response peaks is seen to decrease with increasing $\delta\omega$. In this sense, the optical response of the system is allocated across more frequencies, but with less response at each frequency. In this particular case, the sum rule $\sum_{m=-\infty}^\infty J_m^2 =1$ fixes the total response across all frequencies \cite{clark2019interacting}.

\emph{Bulk wave propagation:} Each of the three examples of optical response regime described above comes with its own implications for wave propagation in a bulk optical system which is characterized by these single-particle models. To demonstrate this, we consider a bulk optical medium which consists of time-dependent point particles packed with a number density $n$. The system is equivalently described by a plasma frequency $\omega_p$. It is then possible to compute the dispersion relation of plane waves which propagate in such a uniform time-dependent medium. Figs. \ref{fig:2LS}d-f show the dispersion relations of bulk media with $\omega_p = \omega_0/2$ for the modulation parameters given in each corresponding column. Dispersion relations indicate the relationship between the wavevector $k$ and the quasienergy $\Omega$, which is taken to lie in the first temporal Brillouin zone (TBZ) $-\Omega_0/2 < \Omega \leq \Omega_0/2$. 
For the undriven medium ($\delta\omega = 0$), the dispersion relation is the same as that of a time-independent Lorentz oscillator, but with frequencies folded into the first TBZ. Features such as the light-like and polariton-like parts of the dispersion can still be identified. In this regime, wavepackets propagate in the usual way (Fig. \ref{fig:2LS}g). 

Stronger driving ($\delta\omega/\omega_0 = 0.4$) brings new changes to the band structure. For example, bands near the edges at $\Omega = \Omega_0/2$ have moved up and down in pairs (marked by an arrow in Fig. \ref{fig:2LS}e), and some curvature has been introduced into bands. Additionally, there is an avoided crossing of the two lowest bands. While the wavepacket at the point marked on the band structure propagates coherently and with a similar group velocity to its unmodulated counterpart, a new beating behavior emerges in the amplitude due to the presence of multiple temporal harmonics in the modes which comprise the packet. As it propagates, the wavepacket exchanges energy back and forth with the medium through this behavior which is only possible in the presence of broken time-translation symmetry. 

With driving strength in the extreme nonperturbative regime ($\delta\omega/\omega_0 = 1.5$), the band structure changes substantially. Most prominently, wavevector gaps are introduced, representing wavelengths which cannot propagate in the medium. Band gaps in ``photonic time crystals'' have been identified previously in non-dispersive settings \cite{biancalana2007dynamics, reyes2015observation}. Additionally, the crossed bands shown in Figs. \ref{fig:2LS}d,e are seen to hybridize with one another, eliminating these sharp crossings. 
The panel below (Fig. \ref{fig:2LS}i) shows that a wavepacket centered around the marked mode propagates with amplitude oscillations, as well as dispersion. This dispersion can be attributed to the band curvature which has developed (red ``x'' in Fig. \ref{fig:2LS}g), as compared to the linear dispersion at the corresponding points of Figs. \ref{fig:2LS}d,e.

\begin{figure}
    \centering
    \includegraphics[width=\linewidth]{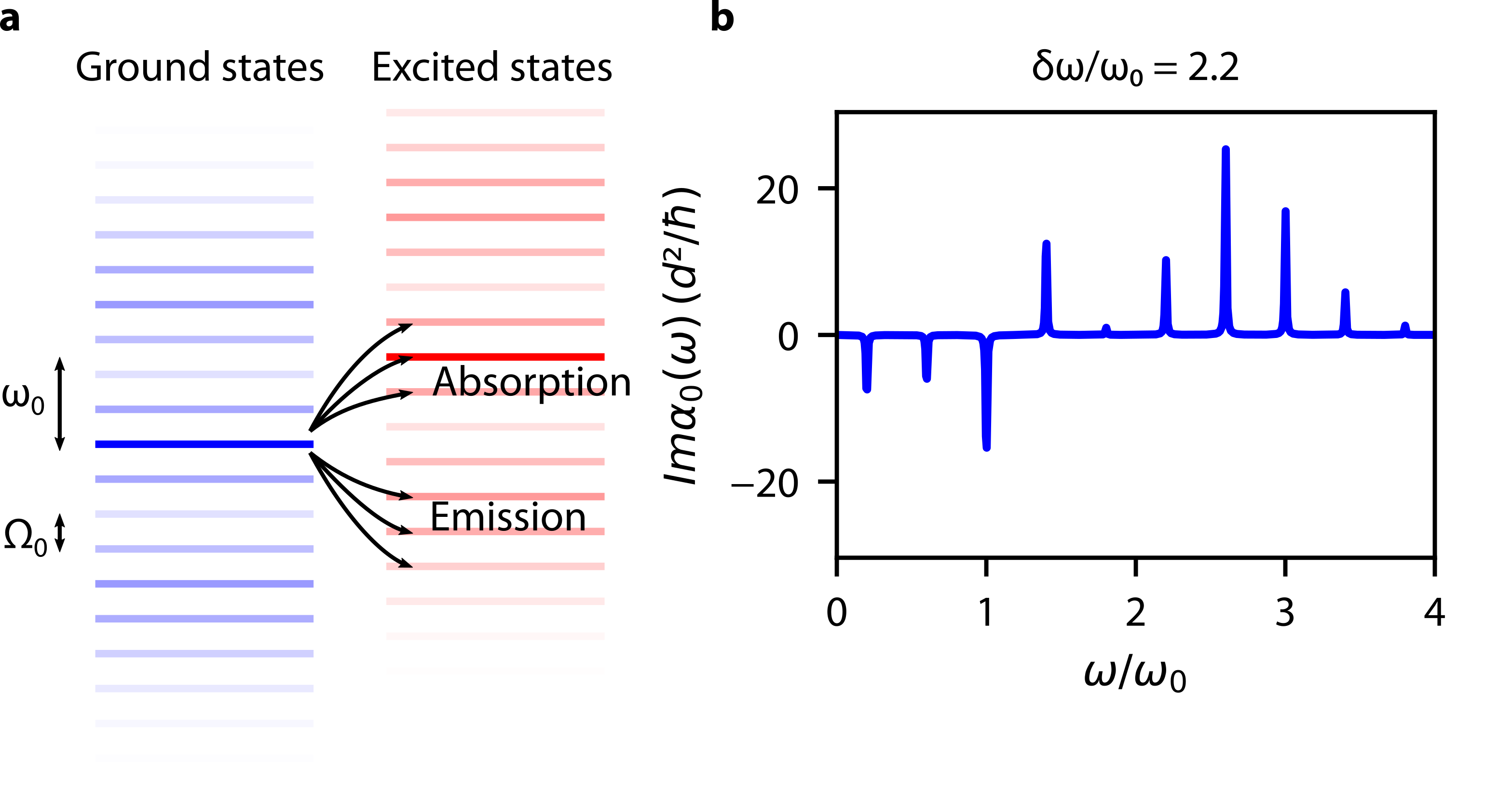}
    \caption{\textbf{Linear response representation of ground state gain in a driven two-level system.} (a) Floquet level diagram for a two-level system driven at frequency $\Omega/\omega_0 = 0.4$ with strength $\delta\omega/\omega_0 = 2.2$. Arrows indicate absorptive and emissive transitions from the thermodynamic ground state to the excited state. (b) Energy transfer properties of the driven system can be visualized through $\Im \alpha_0(\omega)$. Transitions with loss correspond to peaks where $\Im \alpha_0(\omega) > 0$, while transitions with gain correspond to peaks where $\Im \alpha_0(\omega) < 0$.}
    \label{fig:ground_state_gain}
\end{figure}

\textit{Loss and gain:} We now discuss loss and gain in these types of systems, as described in the time-dependent linear response framework. In the absence of any driving, it is well known that a two-level system in its thermodynamic ground state can only absorb energy. The single-particle polarizability in this case is given by a Lorentzian form (Fig. \ref{fig:2LS}a). The quantity $\Im \alpha_0(\omega) \geq 0$ gives the frequency dependent loss, which peaks at $\omega_0$ due to absorptive transitions from the ground to excited state. In a Floquet system, transitions from the ground to excited state can also occur due to absorption of a photon at a frequency $\omega_0 + k\Omega_0$ for some integer $k$. These resonances correspond exactly to the peaks shown in Figs. \ref{fig:2LS}b,c, and also exhibit the property $\Im \alpha_0(\omega) > 0$. As discussed in the theory section, the quantity $\Im \alpha_0(\omega)$ does possess significance for phase-independent energy transfer. From this we see that the example parameters used in Fig. \ref{fig:2LS} give purely absorptive systems. 

To complete our discussion of a modulated 2LS, we give an example of how such a time-dependent two-level system can exhibit both resonant gain and loss in its thermal equilibrium state. To do so, we consider a modulated two-level system with $\Omega = 0.4\omega_0$. When strongly modulated ($\delta\omega/\omega_0 = 2.2$), a substantial contribution emerges from Floquet sidebands which fall below the level of the original ground state. This means that the system can make ground to excited state transitions at frequencies $\omega_0 + k\Omega < 0$ for sufficiently negative integers $k$. These transitions are schematically shown in Fig. \ref{fig:ground_state_gain}a. In the polarizability, these transitions appear as peaks with $\Im \alpha_0(\omega) < 0$ around the relevant resonances, corresponding to energy gain in the Floquet ground state. The gain peaks appear next to other peaks where $\Im \alpha_0(\omega) > 0$, which correspond to absorptive transitions of the form shown in Fig. \ref{fig:2LS}. Thus, a two-level system modulated in this way can provide either absorption or gain to a probe field, depending on the frequency. 

\subsection{Time-dependent Lorentz oscillator}
\label{subsec:LPO}

In this section, we use our framework to describe a medium which behaves as a harmonic oscillator with a time-varying frequency. One system which can be modeled this way under certain conditions is a polar insulator which is strongly driven by an external field. In polar insulators, such as silicon carbide (SiC) and hexagonal boron nitride (hBN), the optical response over some frequency ranges is dominated by optical phonon resonances \cite{basov2016polaritons}. In undriven polar insulators, these resonances lead to well-established peaks at the transverse optical (TO) phonon frequency $\omega_{\text{TO}}$, which are described by a Lorentz oscillator model. However, in the presence of strong laser pulses, the TO phonon frequency of such a material may acquire a time dependence \cite{cartella2018parametric}. If the frequency of the modulating pulse is on the order of $\omega_{\text{TO}}$ itself, then a dispersive time-dependent framework is needed to capture optical behaviors around $\omega_{\text{TO}}$. 

To do this, we use a model which we refer to as the ``Lorentz Parametric Oscillator'' (LPO). In this model, we assume that the response of the polar insulator can be characterized by that of a collection of point-like polarizable particles. Each of these particles is a harmonic oscillator with a time-varying resonance frequency $\omega(t)^2 = \omega_0^2(1 + f(t))$, so that the Hamiltonian governing the oscillator is $H_{\text{LPO}}(t) = \frac{p^2}{2m} + \frac{1}{2}m\omega(t)^2 x^2$, where $x$ and $p$ are the position and momentum operators, and $m$ is the effective mass of the atom in the lattice. For $|f(t)| \ll 1$, the first order perturbative correction to the polarizability is given as 
\begin{equation}
    \alpha(\omega, \omega') = 2\pi\alpha^{(0)}(\omega)\delta(\omega-\omega') + \alpha^{(1)}(\omega, \omega') 
\end{equation}
where $\alpha^{(0)}(\omega) = \frac{q^2}{m} \frac{1}{\omega_0^2 - \omega_0^2 - i\omega\Gamma}$ 
is the ordinary Lorentz Oscillator contribution from $f(t) = 0$. In cases where the perturbative approximation breaks down, one can equivalently solve the equation of motion for a harmonic oscillator with time-varying frequency (sometimes referred to as the ``Mathieu equation'' \cite{ruby1996applications}) numerically and take Fourier transforms in order to obtain $\alpha(\omega, \omega')$ more generally. However, in most cases, the perturbative regime should apply, and the first order frequency space correction is given by
\begin{equation}
    \alpha^{(1)}(\omega, \omega') = -\frac{q^2}{m}\frac{\omega_0^2f(\omega-\omega')}{(\omega_0^2 - \omega^2 - i\omega\Gamma) (\omega_0^2 - \omega'^2 - i\omega'\Gamma)},
    \label{eq:LPO_response}
\end{equation}
where $f(\omega)$ is the Fourier transform of the modulation. The expression features two resonant Lorentz oscillator factors in the denominator at $\omega$ and $\omega'$, and is consistent with expressions for resonant nonlinearities, such as Kerr nonlinearities around an atomic resonance \cite{boyd2019nonlinear}.

\begin{figure}
    \centering
    \includegraphics[width=\linewidth]{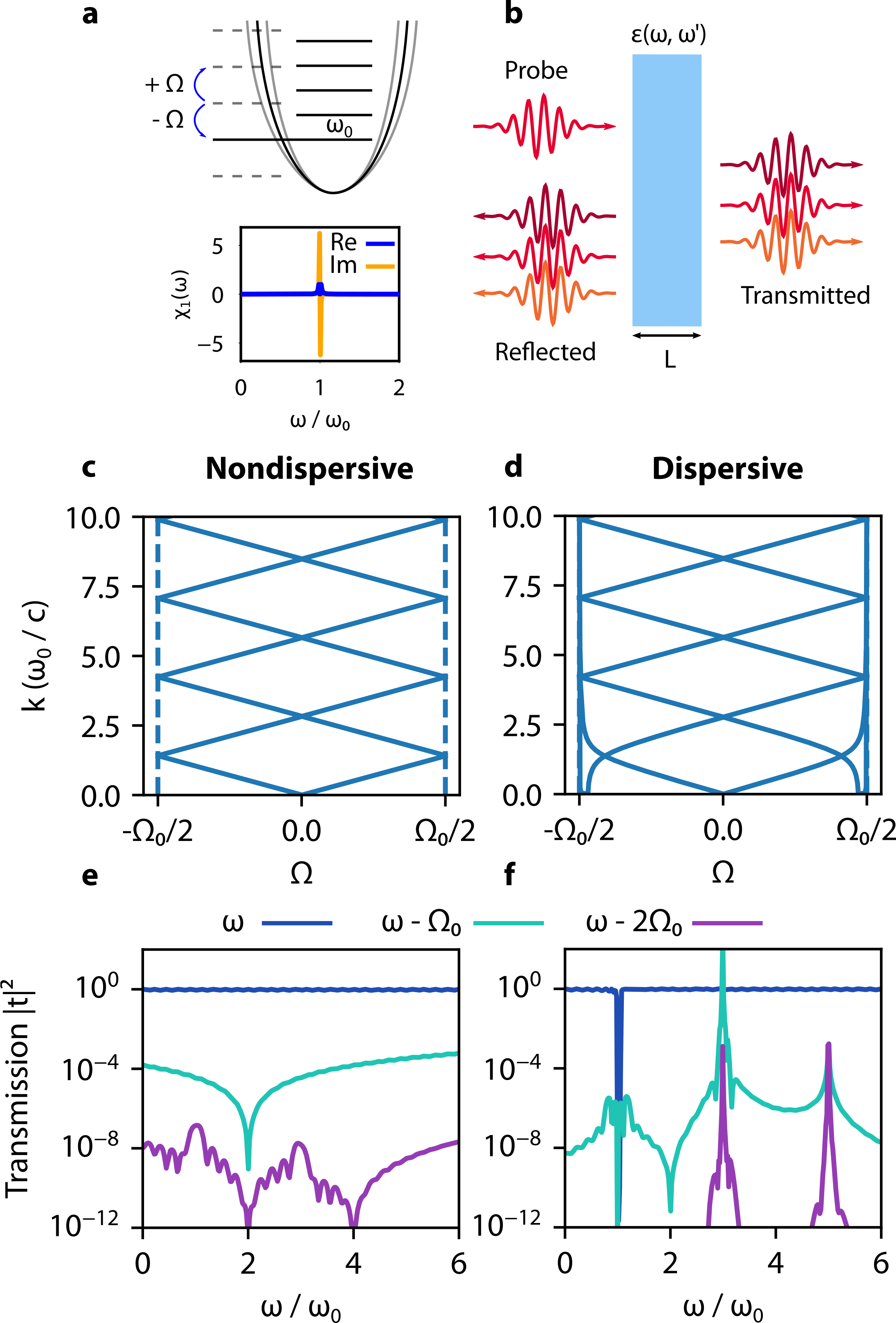}
    \caption{\textbf{Optics of a time-modulated harmonic oscillator.} (a) Quantum harmonic oscillator of frequency $\omega_0$ subject to a frequency-modulation at frequency $\Omega_0$. Panel below shows the real and imaginary parts of $\chi_1(\omega)$ for parametric resonance when $\Omega_0 = 2\omega_0$, at a modulation strength of $\delta\omega  = 10^{-3} \Omega_0$. (b) Incident probe field on a slab of material described by $\varepsilon(\omega, \omega')$. The slab of material has thickness $L$. Since the medium is time-varying, reflected and transmitted waves can be shifted by integer multiples of the drive frequency $\Omega_0$. (c) Band structure of a nondispersive medium with a time-dependent refractive index profile. This is the homogeneous medium dispersion relation $\omega = ck/n$ folded into the temporal Brillouin zone $-\Omega_0/2 < \Omega \leq \Omega_0/2$. (d) Band structure for the dispersive Lorentz parametric oscillator medium depicted in (a), with $\Omega_0 = 2\omega_0$. (e-f) Transmission spectrum for probe and shifted probe frequencies for the configuration depicted in (b) for both nondispersive and dispersive modulations.}
    \label{fig:LPO}
\end{figure}

To give an example of how this dispersive time-dependent response function can be used in optics, we consider the scattering of an incident wave from a slab of material described by $\chi(\omega, \omega')$ for a periodic modulation $f(t) = \delta\omega \cos(\Omega_0 t)$. The general concept of scattering from dispersive time-dependent materials was recently explored in \cite{ptitcyn2021scattering}. In this example, we will focus specifically on how dispersive resonance can greatly impact the reflection and transmission of waves from a material. To demonstrate this, we consider a thin film scattering problem which consists of a weak probe field at frequency $\omega_p$ incident on a material $\varepsilon(\omega, \omega')$ of length $L$. If the material is time dependent with some periodicity $\Omega_0$, then in general, there will be reflected and transmitted waves of shifted frequencies $\omega_p + k\Omega_0$, where $k$ is an integer. To elucidate the effect of dispersive resonances, we compare the scattering problem for two different time-dependent materials: (1) a nondispersive material which has a permittivity $\varepsilon(t) = \varepsilon_{\text{bg}}(t) + \delta\varepsilon\,f(t)$, and (2) a dispersive material described the LPO model detailed above. For both materials, we use a periodic modulation profile $f(t) = \cos(\Omega_0 t)$, specifically focusing on the parametric resonance case given by $\Omega_0 = 2\omega_0$. 

Figs. \ref{fig:LPO}c-d show the photonic time-crystal band structures corresponding to materials (1) and (2) for $\delta\varepsilon = \delta\omega = 10^{-3}$. In the absence of dispersion, these two descriptions coincide. In the nondispersive case, the weak interaction means that the band structure simply corresponds to the undriven material dispersion $\omega_k = ck/\sqrt{\varepsilon_{\text{bg}}}$ folded into the first temporal Brillouin zone with quasifrequencies $-\Omega_0/2 < \Omega \leq \Omega_0/2$. Similarly, the dispersive band structure can be understood as that of an undriven Lorentz Oscillator dispersion folded into the TBZ. In this particular case we have chosen for parametric resonance ($\Omega_0 = 2\omega_0$), the steep branch of the dispersion due to the resonance at $\omega_0$ coincides with the band edge at $\Omega_0/2$. This parametric resonance leads to pronounced effects on incoming waves.

In Figs. \ref{fig:LPO}e-f, we show results for the transmission amplitude for an incident field at $\omega$ for shifted frequencies $\omega$, $\omega - \Omega_0$, and $\Omega_0$ for a thin film created from the nondispersive and dispersive materials described above. For the nondispersive modulation, the vast majority of the transmitted field lies at the incident frequency. In contrast, the dispersive material exhibits peaks of resonant conversion for the incident frequency $3\omega_0$. This occurs because at the parametric resonance condition ($\Omega = 2\omega_0$), the downshifted harmonic lies at $\omega_0$, which is resonant with the oscillator denominators of Eq. \ref{eq:LPO_response}. Since the interaction takes place in a thin film, radiation at incident or new frequencies may continue to re-interact with the material, leading to cascaded harmonics. This is, for example, the origin of a similar response for incident field of $5\omega_0$. The differences between behavior between the dispersive and nondispersive models highlight the importance of using a model which is consistent with underlying microscopic dynamics.

We now comment briefly about the relationship between the time-modulated two-level and Lorentz oscillator models. In the limit of time-independent systems, it is well known that both of these models exhibit the same form of dipole response, given by $\alpha^{(0)}(\omega)$. The intuition behind this is that a weak field which probes the ground state of a harmonic oscillator can effectively only ``see'' the first transition of the oscillator ladder, so the two-level model is recovered. This close relationship dissolves when time-modulations are introduced, as we have seen when comparing the two systems. A parametric drive involves more states of the harmonic oscillator ladder into the dynamics, so that the parametric oscillator model is fundamentally different than a modulated two-level system. The departure of these models here is not dissimilar to what unfolds in the nonlinear response of the static systems: the two-level system displays resonant nonlinearities, while the oscillator exhibits no nonlinearity at all. This serves as a clear example, then, that models for the optical response of strongly driven materials should be considered carefully on the basis of microscopic dynamics.





\begin{figure*}
    \centering
    \includegraphics{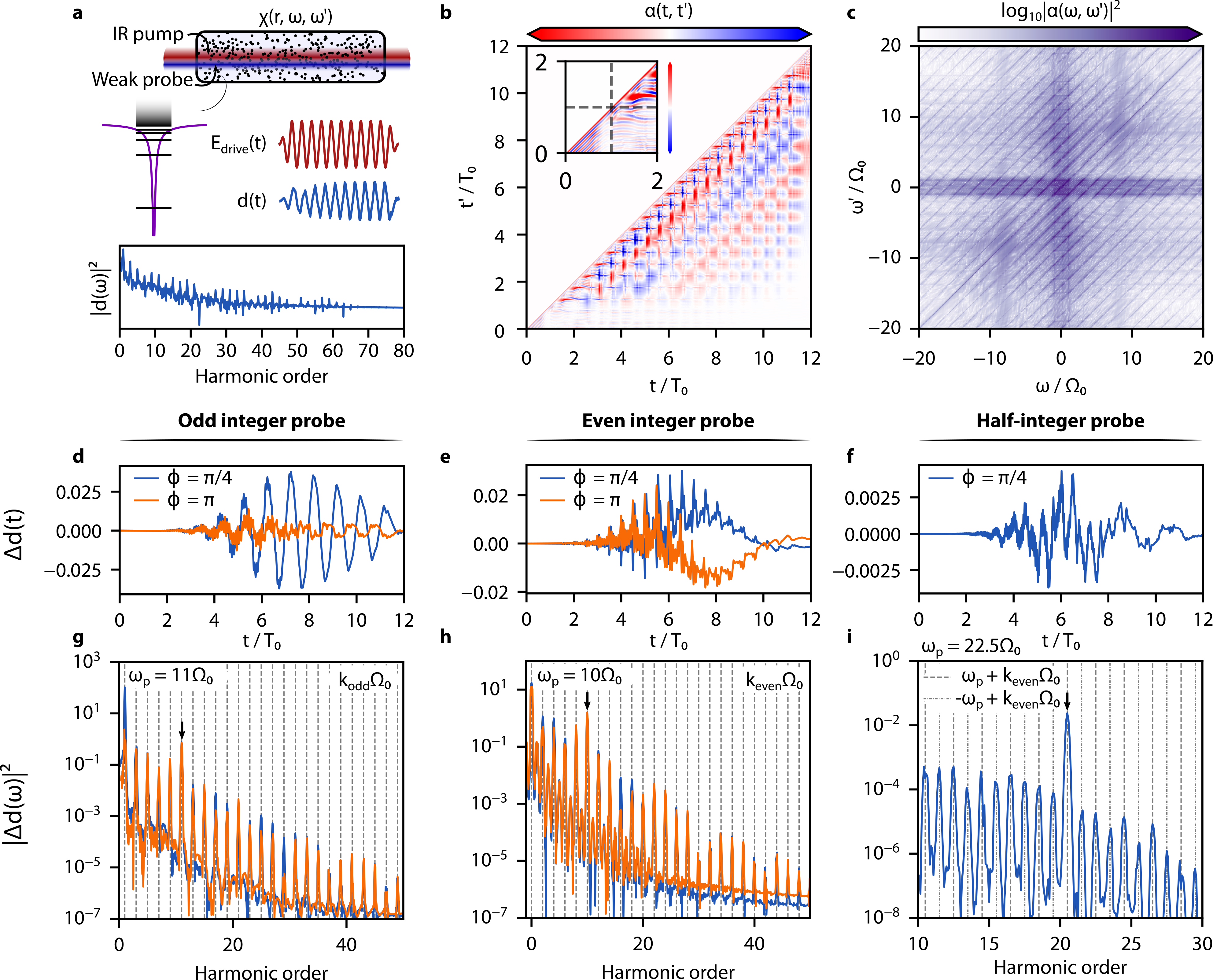}
    \caption{\textbf{High harmonic generation (HHG) as a time-dependent optical medium.} (a) The general setup of HHG consists of a gas sample which is irradiated with an extremely strong IR laser pulses $E_{\text{drive}}(t)$. In a 1D Coulomb potential model, the ground state electron is ejected into the ionized continuum, generating a dipole moment $d(t)$ in the process. This induced dipole moment contains many harmonics of the drive frequency, as shown by the plot of $|d(\omega)|^2$. The harmonics continue up to a cutoff frequency, which is related to the ionization energy of the atom. (b) Two-time atomic susceptibility $\alpha(t, t')$ of a single particle which undergoes the modulation shown in (a). (c) Two-frequency atomic susceptibility $\alpha(\omega, \omega')$ which shows the frequency decomposition of $\alpha(t, t')$. Harmonic stripes can be seen for $\omega' = \omega + k_{\text{even}} \Omega_0$ due to the quasi-periodic nature of the modulation. (d-f) Induced dipole moments by probe fields at different frequencies and phases. (g-i) Induced dipole moment spectra corresponding to each of the probes in (d-f). Probe frequencies are marked with an arrow. Frequencies of the peaks are marked with dashed lines.}
    \label{fig:HHG}
\end{figure*}

\subsection{High harmonic generation}
\label{subsec:HHG}

In this section, we show how the time-dependent linear response framework provides a pathway to describe IR to X-ray frequency optics of systems which exhibit high harmonic generation (HHG). The most basic configuration for such a system is a gas cell which is pumped with an extremely intense infrared laser pulse. For sufficiently strong pumps, the system exhibits highly nonperturbative effects of strong field physics, and many pump photons can be converted into single photons of frequencies which are more than one hundred times higher than that of the pump \cite{mcpherson1987studies, lewenstein1994theory}. These systems serve as valuable sources of UV and X-ray photons which are difficult to produce by any other means, and also generate harmonic combs which form attosecond pulses \cite{paul2001observation, krausz2009attosecond}. More recently, HHG has also been observed from solids \cite{ghimire2019high}.

Although HHG systems have been studied for decades for light generation, there are untapped opportunities to use them as venues for new optical interactions. We propose that HHG systems represent an intriguing platform to study the optics of time-varying materials at high frequencies. From this point of view, the strongly driven gas can itself be considered a time-varying optical medium. Due to the strong field strengths and atomic resonances involved in these systems, dispersion plays an important role.     

The general setup of an HHG system is shown in Fig. \ref{fig:HHG}a. In the absence of any probe field, the driven system acquires a dipole moment which oscillates at many harmonic multiples of the driving frequency, leading to HHG. If the emitted photons are considered quantum mechanically, HHG can be equivalently characterized as spontaneous emission from transitions between the Floquet quasi-energy levels of the driven system \cite{gorlach2020quantum}.

Using a 1D model of an atomic potential, we numerically solved the time-dependent Schrodinger equation for $H(t) = \frac{p^2}{2m} + V(x) - E_0 x \sin(\Omega_0 t) g(t)$, where $V(x) = -1/\sqrt{x^2 + a^2}$ is a ``soft Coulomb'' potential regulated by the parameter $a$, and $g(t)$ is an envelope function which turns the drive on and off. Parameters were chosen so that the ionization energy of the potential matches that of Neon. Using numerical evolution of this Hamiltonian, we directly computed the atomic polarizability $\alpha(t, t')$ using the Kubo formula (Eq. \ref{eq:time_domain_kubo}) \footnote{In a Hermitian system, the polarizability $\alpha(t,t')$ can be computed by evolving all eigenstates of the Hamiltonian in time, and then taking expectation values of the appropriate operators to implement the time-domain Kubo formula directly. However, the numerical models used for HHG typically rely on absorbing boundary conditions, which break the Hermiticity, and thus the completeness of eigenstates. Therefore, we compute the time-domain response function via the Liouvillian evolution in a manner consistent with the quantum regression theorem. We have verified that the response function $\alpha(t,t')$ obtained through this method generates dipole responses to probe fields which match those obtained by directly incorporating the probe into the Hamiltonian.}. The results of this calculation are shown in Fig. \ref{fig:HHG}b. As dictated by the causality constraint, the dipole response is nonzero only for times $t > t'$.
Even though this system is driven for a relatively small number of periods, some features of the Floquet regime emerge. From the theoretical discussion around Eq. \ref{eq:harmonic_response_function}, we know that for a perfectly time periodic system, the frequency polarizability $\alpha(\omega, \omega')$ converges to a series of delta functions spaced at integer multiples of the drive. By numerically transforming the time-domain polarizability, we show that this holds approximately true. The squared magnitude of the frequency-domain polarizability $|\alpha(\omega, \omega')|^2$ is shown in Fig. \ref{fig:HHG}c over a range of harmonics. The clear diagonal stripes adhere to lines for which $\omega' = \omega + k\Omega_0$. For a general system, $k$ can be any integer. However, the inversion symmetry of the potential and driving field we have chosen dictates that $k$ may only take on even values, as is consistent with the general framework for selection rules in HHG \cite{neufeld2019floquet}. In the limit of weak driving, where only a small number of harmonics can be produced, this constraint reduces to the well-known fact that centrosymmetric materials have no $\chi^{(2)}$ nonlinearity. 

We now explore the consequences of these response functions for weak probes which interact with the driven system. Figs. \ref{fig:HHG}d-f, show the change to the time-dependent dipole moment $\Delta d(t)$ induced by a probe field $E(t) = E_p \cos(\omega_p t - \phi) e^{-(t - t_0)^2/2\tau^2}$ for different probe frequencies and phases. Generically, the dipole moment peaks can appear at $\pm\omega_p + k\Omega_0$, where $k$ is an even integer. Different behaviors emerge depending on the frequency $\omega_p$ and phase $\phi$ of the probe, which are visualized through the frequency spectrum $|\Delta d(\omega)|^2$ for different probe parameters.

For an odd-harmonic probe ($\omega_p = 11\Omega_0$), the dipole responds at other odd frequencies (Fig. \ref{fig:HHG}g). A black arrow marks harmonic 11, which oscillates at an amplitude higher than that of the surrounding peaks. Nevertheless, notable contributions come from many peaks, extending up through around harmonic 50. Additionally, we note that the phase of the probe with respect to the pump can contribute substantially to the resulting output. In particular, we show examples of the phases $\phi = \pi/2, \pi$. For some of the harmonics produced by the probe, the amplitude can vary by an order of magnitude or more depending on the probe phase. This type of behavior has actually already been observed in the context of so-called ``two-color high harmonic generation'' in which some harmonic of the drive (usually the third), is sent into the sample along with the drive itself \cite{kim2005highly}. In some sense, the pump-probe schemes discussed here are similar in nature. The main development here is that while experiments so far have considered only a few harmonics of the drive as a probe, our insights from Floquet linear response indicate that such HHG systems should also exhibit response at many harmonics of the drive, providing a way to engineer optical response at UV and X-ray frequencies.

So far, we have shown that an odd harmonic probe can essentially modify the normal HHG spectrum (which also consists only of odd harmonics in this example). However, we now show that weak probe fields sent at the time-driven system can also be used to generate dipole moments at frequencies which are not produced in absence of the probe. For example, sending in a probe frequency at some even harmonic $m$ induces a comb of dipole moments at frequencies $k_{\text{even}}\Omega_0$. This change to the dipole moment will radiate at even harmonics, which are not produced by the system in absence of the probe. Such a configuration is shown in Figs. \ref{fig:HHG}e,h, where the probe is sent at harmonic 10. The result is an even comb of induced dipole moments, which will then radiate into even harmonics. Similarly to the odd-probe, the probe harmonic stands out in strength above the others, and the probe phase can strongly influence the output.

Finally, we show that by probing at a non-harmonic frequency results in a pair of interleaved combs (Fig. \ref{fig:HHG}f,i). In particular, sending in a weak probe of harmonic $\omega_p = 22.5\omega_0$ produces dipole moment peaks at $\pm \omega_p + k_{\text{even}} \Omega_0$, leading to induced dipole moments at non-harmonic frequencies, but which are separated by even multiples of the drive. Moreover, this example indicates that the driven system will respond optically at tens of harmonics, which for a near-IR pump corresponds to optical response at wavelengths of 10's of nanometers or below. Thus, these results show promise for the potential to use existing HHG systems as a platform to study the optical response of strongly driven systems, potentially leading to controllable optical materials which can respond and convert frequencies in the UV and X-ray regime. While we have focused for clarity on a single-particle polarizability model, the two-frequency linear response framework naturally lends itself to the inclusion of spatial aspects of HHG problems, which can enable studies of phase-matching and wave propagation \cite{l1991higher, salieres1995coherence}. 

We additionally note that the time-dependent linear response is particularly appealing for the study of probe-response in HHG, because once a function such as $\alpha(t, t')$ is computed through potentially time-consuming quantum mechanical simulations (i.e. results of Fig. \ref{fig:HHG}b), the response to any probe can be computed as a simple convolution integral. In fact, through energy dissipation/gain measurements, it may be possible in some cases to determine $\alpha(\omega, \omega')$ experimentally, enabling inferences about how the system will response to other probes. Such an approach may be particularly appealing in order to construct probe fields which will selectively enhance or suppress the generation of certain harmonics, which could relate to optimization of 2-color HHG processes \cite{kim2005highly}, or more complex analogs. This framework could also be used to describe the high frequency parametric gain which has been observed in HHG systems \cite{seres2010laser}, allowing for the creation of more efficient systems which amplify high frequency radiation.


\section{Conclusion and outlook}

In summary, we have presented a framework for describing electromagnetic response and wave propagation in dispersive time-varying quantum systems. We have established fundamental properties of time-varying response functions, with a special focus on developing forms for use in time-periodic (Floquet) systems. We have addressed fundamental questions about the nature of energy transfer (gain and loss) in these systems. In fact, the relationship between K.K. relations and energy transfer in time-varying materials that we have established can enable the use of absorption/gain measurements to construct the full complex response functions of time-varying systems. Additionally, we have shown selected examples of this framework to address a diverse set of problems which raise implications for superconducting microwave circuits, polar insulators in the IR, and UV/X-ray optics of HHG systems. We hope that this unifying approach will reveal further similarities among fields which would normally be considered disparate. 

One important future direction is the development of microscopic models to describe time-varying linear response in more complex systems. For example, many recent works have focused on the electronic states that can be created in Floquet-driven matter (with a particular focus on topological electronic properties) \cite{bao2022light}. However, there is still much work to be done to use these descriptions of strongly driven solids to infer the optical properties of such materials. In some cases, free electron or few-band models may be sufficient to capture the key physics. In more complicated situations, time-dependent density functional theory (TDDFT) may serve as an essential tool for computing time-varying response functions (for example, $\varepsilon(\omega, \omega')$ for a strongly driven semiconductor). Once the optical response of a strongly driven material is appropriately characterized, it can be incorporated into either classical or quantum descriptions of electromagnetic phenomena.

In the classical domain, time-varying materials can lead to the propagation of new types of excitations in materials, and new mechanisms for gain. It is well known that the propagation of waves in a medium with a simple periodic time-varying permittivity can in theory lead to PTCs with momentum bandgaps. In more complex settings where dispersion is important, strong temporal driving may lead to the generation of new ``floquet polaritons'' in either bulk or structured media. Such floquet polaritons on 2D materials could open up a new set of directions for the broad field of polaritonics. 


In the quantum domain, the appropriate use of response functions to describe time-varying materials may enable a general description of quantum light-matter interactions in time-varying materials. 
This can eventually lead to an accurate quantum picture of ``photonic quasiparticles'' \cite{rivera2020light} in time-varying materials which may interact with matter. The study of fundamental quantum light-matter interaction processes in time-driven materials is the first step toward answering questions about how they can be used to construct new devices such as amplifiers or lasers with new output properties, or at frequencies which have been historically difficult to achieve.


\begin{acknowledgments}
    We thank Prof. Ido Kaminer for useful discussions. J.S. acknowledges support a Mathworks fellowship, as well as earlier support from a National Defense Science and Engineering Graduate Fellowship of the Department of Defense (F-1730184536). N.R. acknowledges the support of a Junior Fellowship from the Harvard Society of Fellows, as well as earlier support from a Computational Science Graduate Fellowship of the Department of Energy (DE-FG02-97ER25308), and a Dean's Fellowship from the MIT School of Science. This material is based on work supported in part by the Defense Advanced Research Projects Agency (DARPA) under agreement no. HR00112090081. This material is based upon work supported in part by the Air Force Office of Scientific Research under the award number FA9550-20-1-0115; the work is also supported in part by the U. S. Army Research Office through the Institute for Soldier Nanotechnologies at MIT, under Collaborative Agreement Number W911NF-18-2-0048. 
\end{acknowledgments}

\bibliographystyle{ieeetr}
\bibliography{vacuum.bib}

\newpage
\appendix

\onecolumngrid

\section{Derivations of general properties}
\subsection{Kubo formula for Floquet systems}

Here, we derive an expression for the two-frequency atomic polarizability $\alpha(\omega, \omega')$ of a generic single-electron system which is varied in time. We assume that the unperturbed system is described by a Hamiltonian $H_0(t)$, with the corresponding time evolution operator $U_0(t)$ which satisfies $i\hbar\partial_t U_0(t) = H_0(t)U_0(t)$  We can send an electric field probe which couples to the dipole operator $d$ of the unperturbed system as $V(t) = -d E(t)$. We have written our expressions in terms of a single coordinate, but all that follows can be easily generalized to include vector directions for fields and dipole moments. Then, by linear response theory, the change in the dipole moment is given as 
\begin{equation}
    \delta\Braket{d(t)} = \int_{-\infty}^\infty dt'\,\alpha(t, t') E(t'),
\end{equation}
where the Kubo formula for the two-time polarizability is given as 
\begin{equation}
    \alpha(t,t') = \frac{i}{\hbar}\theta(t-t')\Braket{\psi_0|[d_I(t), d_I(t')]|\psi_0}.
\end{equation}
Here, $d_I(t) = U_0^\dagger(t) d U_0(t)$ is the dipole moment operator in the interaction picture, and $\ket{\psi_0}$ is the state of the system before the probe is applied. In the frequency domain, we can write
\begin{equation}
    \delta\Braket{d(\omega)} = \int_{-\infty}^\infty \frac{d\omega'}{2\pi}\alpha(\omega, \omega') E(\omega'),
\end{equation}
where $\alpha(\omega, \omega') \equiv \int dt\,dt'\, e^{i(\omega t - \omega't')} \alpha(t,t')$. In general, $\alpha(\omega, \omega')$ can be a function of the two continuous frequencies $\omega, \omega'$. 

We may also specialize this result to systems which are time-modulated periodically. This will allow us to make key simplifications. Namely, Floquet theory will be used to decompose the problem into harmonics. In this case, we assume that the time dependent Hamiltonian $H_0(t)$ has a period $T$, so that $H_0(t + T) = H_0(t)$. The frequency associated with this period is $\Omega_0 \equiv 2\pi/T$. In this case, solutions to the time-dependent Schrodinger equation $i\hbar\partial_t\ket{\psi_\alpha(t)} = H_0(t)\ket{\psi_\alpha(t)}$ can be written in terms of Floquet states
\begin{equation}
    \ket{\psi_\alpha(t)} = e^{-i\epsilon_\alpha t} \ket{\phi_\alpha(t)}
\end{equation}
where $\epsilon_\alpha$ is the Floquet quasi-energy which lies in the Brillouin zone, and $\ket{\phi_\alpha(t)}$ is a periodic function (known as a Floquet mode) which can be decomposed in terms of harmonics as $\ket{\phi_\alpha(t)} = \sum_n e^{in\Omega_0 t}\ket{\phi_\alpha^n}$

By assuming periodicity, we can write a form for $\alpha(t,t')$ in terms of the Floquet modes. Inserting a complete set of Floquet states $\ket{\psi_\alpha(0)}$, we can write
\begin{align}
    \Braket{\psi_0|d_I(t)d_I(t')|\psi_0} &= \sum_\alpha \Braket{\psi_0|U_0^\dagger(t) d U_0(t) |\psi_\alpha} \Braket{\psi_\alpha|U_0^\dagger(t') d U_0(t') |\psi_0} \\
    &= \sum_\alpha \Braket{\psi_0(t)|d|\psi_\alpha(t)}\Braket{\psi_\alpha(t')|d|\psi_0(t')}
\end{align}
Substituting these states into the above expression for dipole expectation value, we find
\begin{equation}
    \Braket{\psi_0|d_I(t)d_I(t')|\psi_0} = \sum_\alpha \sum_{k,l,m,n} e^{-i(\epsilon_\alpha - \epsilon_0)t + i(n-k)\Omega_0 t} e^{i(\epsilon_\alpha-\epsilon_0)t' - i(m-l)\Omega_0 t'} \Braket{\phi_0^k|d|\phi_\alpha^n}\Braket{\phi_\alpha^m|d|\phi_0^l}
\end{equation}
Next we use $\theta(t-t') = i\int\frac{d\omega''}{2\pi}\frac{e^{-i\omega''(t-t')}}{\omega'' + i\eta}$ to write 
\begin{equation}
    \begin{split}
        \frac{i}{\hbar} \int dt\,dt'\,e^{i\omega t - i\omega't'}\theta(t-t')& \Braket{\psi_0|d_I(t)d_I(t')|\psi_0} 
        = -\frac{1}{\hbar} \int\frac{d\omega''}{2\pi}\frac{1}{\omega'' + i\eta}\int dt\,dt'\,e^{i\omega t - i\omega't'} e^{-i\omega''(t-t')} \\
        &\times\left(\sum_\alpha \sum_{k,l,m,n} e^{-i(\epsilon_\alpha - \epsilon_0)t + i(n-k)\Omega_0 t} e^{i(\epsilon_\alpha-\epsilon_0)t' - i(m-l)\Omega_0 t'} \Braket{\phi_0^k|d|\phi_\alpha^n}\Braket{\phi_\alpha^m|d|\phi_0^l}\right)
    \end{split}
\end{equation}
After simplifying this expression, one obtains
\begin{equation}
\begin{split}
    \alpha(\omega, \omega') = -\frac{2\pi}{\hbar}\sum_\alpha \sum_{k,l,m,n} &\left(\frac{\Braket{\phi_0^k|d|\phi_\alpha^n}\Braket{\phi_\alpha^m|d|\phi_0^l}}{\omega + (n-k)\Omega_0 - (\epsilon_\alpha - \epsilon_0) + i\eta} - \frac{\Braket{\phi_0^k|d|\phi_\alpha^n}\Braket{\phi_\alpha^m|d|\phi_0^l}}{\omega - (m-l)\Omega_0 + (\epsilon_\alpha - \epsilon_0) + i\eta}\right) \\
    &\hspace{2cm}\times \delta\left[\omega - \omega' - (m+k-n-l)\Omega_0\right]
\end{split}   
\label{eq:floquet_kubo}
\end{equation}
We immediately note that $\alpha(\omega, \omega')$ takes the form of a sum over delta functions of the form $\delta(\omega-\omega'-k\Omega_0)$, where $k$ is an integer. In other words, an incident field at frequency $\omega'$ can induce a dipole moment at frequencies $\omega' +k\Omega_0$. This property emerges solely from assuming that the system is periodic. As such, for periodic systems, we may replace $\alpha(\omega, \omega')$ with a series of single-argument functions $\alpha_k(\omega)$ defined by
\begin{equation}
    \alpha(\omega, \omega') = \sum_k \alpha_k(\omega) 2\pi\delta(\omega-\omega'-k\Omega_0).
\end{equation}

\subsection{Derivation of generalized K.K. Relations}
\label{subsec:kk_appendix}

In this section, we derive Kramers Kronig (K.K.) relations for linear response functions $\chi(\omega, \omega')$ which are not time-translation invariant. The conventional proof of Kramer's Kronig relations uses complex analysis to show that optical passivity of a linear response function $\chi(\omega)$ implies to the relation
\begin{equation}
    \chi(\omega) = \frac{i}{\pi}\mathcal{P}\int d\omega''\frac{\chi(\omega'')}{\omega - \omega''},
    \label{eq:conventional_kk}
\end{equation}
where $\mathcal{P}$ denotes the principle value of the integral. Then, splitting this equation into complex components yields a direct relationship between the real and imaginary parts of $\chi(\omega)$. In a passive system, $\Im \chi(\omega)$ encodes the dissipation of the material. A basic consideration of energy conservation requires that $\Im \chi(\omega) \geq 0$ for all $\omega > 0$ to ensure that inputs are attenuated over time, rather than amplified. While the conventional K.K. relation is usually explained in terms of optical passivity (and complex poles in the upper half plane), Eq. \ref{eq:conventional_kk} can also be derived as an immediate consequence of \emph{causality}: the fact that a system can only respond to an impulse after its application.

While linear response functions in time-varying systems are not necessarily passive, they must respect causality. We will use this constraint to derive K.K. relations for linear response functions in time-dependent systems. For a time-dependent response function $\chi(t,t')$, the causality constraint can be expressed in terms of the heaviside function $\theta(t)$ as
\begin{equation}
    \chi(t, t') = \theta(t -t')\chi(t, t').
    \label{eq:two_time_causaility}
\end{equation}
To derive the K.K. relation, we take the two-time Fourier transform of both sides, using the convention $\chi(\omega, \omega') = \int dt\,dt'\,e^{i(\omega t - \omega't')}\chi(t,t')$. This allows us to write
\begin{equation}
    \chi(\omega, \omega') = \int dt\,dt'\, e^{i(\omega t - \omega't')} \theta(t-t')\chi(t,t').
    \label{eq:two_time_causality_fourier}
\end{equation}
The right hand side can be evaluated using the convolution theorem. To perform this, we note that the double Fourier transform of the heaviside function is given by
\begin{equation}
    \theta(\omega, \omega') = \int dt\,dt'\,e^{i(\omega t - \omega't')}\theta(t-t') = 2\pi\delta(\omega-\omega')\left[2\pi\delta(\omega+\omega') + \frac{2i}{\omega+\omega'}\right].
\end{equation}
The convolution integral thus gives
\begin{equation}
\begin{split}
    \int dt\,dt'\, e^{i(\omega t - \omega't')} \theta(t-t')\chi(t,t') &= \int \frac{d\nu\,d\nu'}{(2\pi)^2} \chi(\omega - \nu, \omega' - \nu')\theta(\nu, \nu') \\
    &= \frac{1}{2\pi}\int d\nu \chi(\omega - \nu, \omega' - \nu)\left[2\pi(\nu + \nu') + \frac{i}{\nu}\right] \\
    &= \frac{1}{2}\chi(\omega, \omega') + \frac{i}{2\pi}\int d\nu\, \frac{\chi(\omega - \nu, \omega' - \nu)}{\nu}
\end{split}
\end{equation}
Plugging this back into Eq. \ref{eq:two_time_causality_fourier}, we can solve for $\chi(\omega, \omega')$ to give the K.K. relation
\begin{equation}
    \chi(\omega, \omega') = \frac{i}{\pi}\mathcal{P}\int d\omega'' \frac{\chi(\omega - \omega'', \omega' - \omega'')}{\omega''}.
    \label{eq:new_kk}
\end{equation}
To see how this general relation relates to the usual time-translation-invariant case, we take $\chi(\omega, \omega') = 2\pi\delta(\omega - \omega')\chi(\omega)$. Substituting this form into the new relation gives $\chi(\omega) = \frac{i}{\pi}\int d\omega'' \frac{\chi(\omega - \omega'')}{\omega''}$, which after changing the integration variable matches the usual form noted in Eq. \ref{eq:conventional_kk}. The new K.K. given in Eq. \ref{eq:new_kk} importantly indicates that even without passivity, causality still requires a strict relation between the real and imaginary parts of $\chi(\omega, \omega'$. Specifically, these are:
\begin{align}
    \Re \chi(\omega, \omega') &= -\frac{1}{\pi} \mathcal{P}\int d\omega'' \frac{\Im\chi(\omega - \omega'', \omega' - \omega'')}{\omega''} \\
    \Im \chi(\omega, \omega') &= \frac{1}{\pi} \mathcal{P}\int d\omega'' \frac{\Re\chi(\omega - \omega'', \omega' - \omega'')}{\omega''}.
\end{align}

\emph{Specialization to the Floquet case:} While the K.K. relation Eq. \ref{eq:new_kk} is valid for any linear time-varying system, we can make simplifications in the case where the time-varying system is periodic with frequency $\Omega_0$. In this case, we have seen that the response function necessarily takes the form of a sum over harmonics $\chi(\omega, \omega') = \sum_k \chi_k(\omega) 2\pi\delta(\omega-\omega'-k\Omega_0)$. Plugging this assumption into Eq. \ref{eq:new_kk} shows that the harmonics behave independently from one another. Specifically, each harmonic component $\chi_k(\omega)$ actually satisfies the time-invariant K.K. relation 
\begin{equation}
    \chi_k(\omega) = \frac{i}{\pi}\mathcal{P}\int d\omega''\frac{\chi_k(\omega'')}{\omega - \omega''}.
\end{equation}

\section{Two level system derivations}

In this section, we provide details of the derivation of the polarizability for the time-modulated two-level system (2LS) discussed in the main text. There, we assumed that the Hamiltonian of the driven two-level system takes the form
\begin{equation}
    H_0(t) = \frac{\sigma_z}{2}\left(\omega_0 + \delta\omega\cos \Omega_0 t\right).
    \label{eq:2LS_hamiltonian}
\end{equation}
Our goal here is to compute the atomic polarizability, which in time domain is given via the Kubo formalism as 
\begin{equation}
    \alpha(t,t') = \frac{i}{\hbar}\theta(t-t')\braket{\psi_0|[d_I(t), d_I(t')]|\psi_0},
    \label{eq:kubo_time_domain}
\end{equation}
where $d_I(t)$ is the dipole moment operator in the interaction picture of the time dependent Hamiltonian in absence of the probe field. Computing this dipole operator is done with the aid of the unitary time evolution operator $U_0(t)$ which corresponds to $H_0(t)$. This Hamiltonian commutes with itself at all times, which means that the unitary evolution operator can be evaluated directly without any concerns of time ordering as 
\begin{equation}
    U_0(t) = \exp\left[-i\int^t dt'\,H_0(t')\right] = e^{-i\omega_0 \sigma_z t/2} e^{-i(\delta\omega/2\Omega_0)\sigma_z \sin\Omega_0 t}
    \label{eq:2LS_unitary}
\end{equation}

We note that the second can be expanded as a Floquet series using the Jacobi-Anger expansion $e^{iz\sin\theta} = \sum_n J_n(z) e^{in\theta}$. This means that we can write expressions for the time-dependent Floquet states
\begin{align}
    \ket{g(t)} &= e^{-i\omega_0 t/2} \sum_m e^{im\Omega_0 t} J_m\left(-\frac{\delta\omega}{2\Omega_0}\right)\ket{g} \\
    \ket{e(t)} &= e^{i\omega_0 t/2} \sum_m e^{im\Omega_0 t} J_m\left(\frac{\delta\omega}{2\Omega_0}\right)\ket{e}
\end{align}
The states clearly evolve in the form of Eq. A4. We see that the two Floquet states consist of the original eigenstates $\ket{g}$ and $\ket{e}$ with a time-dependent phase attached. This is a feature of this particularly simple example, as the Hamiltonian (Eq. \ref{eq:2LS_hamiltonian}) has only a $\sigma_z$ term, so the states necessarily evolve independently. Thus in this case, the Floquet levels can be written as $\ket{g_m} = J_m(-\delta\omega/2\Omega_0)\ket{g}$ and $\ket{e_m} = J_m(\delta\omega/2\Omega_0)\ket{e}$. 

\subsection{Derivation of the polarizability}

Here, we calculate the interaction picture dipole operator $d_I(t)$ and evaluate $\alpha(t,t')$ directly from the commutator form of the Kubo formula (Eq. \ref{eq:kubo_time_domain}). The dipole operator is proportional to $\sigma_x = \sigma_+ + \sigma_-$, where $\sigma_\pm$ are the standard state raising and lowering operators. Using the unitary time evolution operator (Eq. \ref{eq:2LS_unitary}), we compute the interaction picture operators
\begin{align}
    \sigma_-(t) &= \sigma_- \sum_n J_n e^{-i\omega_n t} \\
    \sigma_+(t) &= \sigma_+ \sum_n J_n e^{i\omega_n t},
\end{align}
where $\omega_n \equiv \omega_0 + n\Omega_0$ is the bare transition frequency shifted by $n$ harmonics of the drive, and we have introduced the notation $J_n \equiv J_n(\delta\omega/\Omega_0)$ for convenience. The relevant commutator for use in the Kubo formula is then easily computed as 
\begin{equation}
    [\sigma_x(t), \sigma_x(t')] = \sigma_z \sum_{m,n} J_m J_n \left(e^{-i\omega_n t} e^{-i\omega_m t} - \mathrm{c.c}\right),
\end{equation}
so that the time domain response function is
\begin{equation}
    \alpha(t, t') = i\frac{d^2}{\hbar}\theta(t-t') \Braket{\sigma_z}_0 \sum_{m,n} \left(e^{-i\omega_n t} e^{-i\omega_m t} - \mathrm{c.c}\right).
\end{equation}
Here, $\Braket{\sigma_z}_0$ is the expectation value of $\sigma_z$ taken in the steady state of the driven system. The exact ground and excited state probabilities will depend on the damping environment of the time-dependent particle, as will be discussed in the next section. Taking two Fourier transforms as defined by Eq. 2 yields the frequency domain result
\begin{equation}
    \alpha(\omega, \omega') = \frac{2\pi d^2}{\hbar} \Braket{\sigma_z}_0 \sum_{m,n} J_m J_n \left[\frac{\delta(\omega - \omega' + (n-m)\Omega_0)}{\omega' + \omega_m + i\Gamma_m} - \frac{\delta(\omega - \omega' - (n-m)\Omega_0)}{\omega' - \omega_m + i\Gamma_m}\right].
\end{equation}
As required by the periodic nature of the problem, the two-frequency polarizability can be written as a sum over delta functions (Eq. A10). In particular, the integer-order polarizability functions can be written as 
\begin{equation}
    \alpha_k(\omega) = \frac{d^2}{\hbar}\Braket{\sigma_z}_0 \sum_n \left[\frac{J_{n+k} J_n}{\omega - \omega_n + i\Gamma_n} - \frac{J_{n-k} J_n}{\omega + \omega_n + i\Gamma_n}\right]
\end{equation}
This result shows that the polarizability of our example time-driven two-level model essentially amounts to the polarizability of many two-level systems at frequencies spaced out from $\omega_0$ by integer shifts of the driving frequency $\Omega$. This result is entirely consistent with the wisdom of Floquet theory which states that a periodically driven quantum system should behave in many ways like a time-independent system with many more quasi-energy levels corresponding to harmonics. As we will detail in the next section, the damping coefficients are given by $\Gamma_n = 2\pi g^2 J_n^2 \rho(\omega_n)$, where $\rho(\omega)$ is the density of electromagnetic states at frequency $\omega$, and $g$ is the weak coupling associated with the undriven system which gives rise to a damping rate $\Gamma_0 \equiv 2\pi g^2 \rho(\omega_0)$.

\subsection{The effects of damping}
\renewcommand{\sp}{\sigma_+}
\newcommand{\sm}{\sigma_-}
\newcommand{\nth}{n_{\text{th}}}

In the section above, we defined the linear response functions $\alpha_k(\omega)$ for a time-modulated two-level system. In doing so, we introduced damping in a somewhat ad-hoc way. This approach is well-established for static systems. In this section, we place these decay coefficients on more rigorous footing by applying standard density matrix damping theory to the Floquet system. We begin by assuming that the driven two-level system, described by $H_0(t)$, is coupled to a continuous bath of harmonic oscillator modes:
\begin{equation}
    H/\hbar = H_0(t) + \sum_k \nu_k b_k^\dagger b_k + \sum_k g_k \sigma_x (b_k + b_k^\dagger).
\end{equation}
Here, $b_k$ is the annihilation operator for the bath mode of frequency $\nu_k$, which is coupled to the dipole of the driven 2LS with a coefficient $g_k$. Furthermore, we assume that the bath modes are in thermal equilibrium such that the reservoir average is $\Braket{b_k^\dagger b_{k'}} = \nth \delta_{kk'}$, where $\nth$ is the photon number of the bath at thermal equilibrium. Following the usual sequence of manipulations \cite{scully1999quantum}, one can find an equation for the reduced density matrix (corresponding to the 2LS only), which is given as:
\begin{equation}
    \dot{\rho}_S(t) = -\mathrm{Tr}_R \int_0^t dt'\, \left[V(t), [V(t'), \rho_S(t')\otimes\rho_R(0)]\right].
    \label{eq:reduced_dm_iterative}
\end{equation}
Here, $V(t)$ is the interaction picture Hamiltonian obtained by transforming the dipole-bath interaction term according to the unitary evolution operator of the non-interaction systems. Note that we have neglected a term which is linear in the bath operators, since such a term will vanish when reservoir expectation values are taken. Additionally, we have made the usual assumption that the density matrix of the reservoir is unchanged by weak interactions with the system, so that the quantity $\rho_R(0)$ can be used at all times. 

For the specific 2LS model that we consider, the interaction picture Hamiltonian is found to be:
\begin{equation}
    V(t) = \sum_k g_k \sum_n J_n \left(\sigma_+ e^{i\omega_n t} + \sigma_- e^{-i\omega_n t}\right)\left(b_k e^{-i\nu_k t} + b_k^\dagger e^{i\nu_k t}\right)
\end{equation}
We can now implement a type of rotating wave approximation (RWA) for each of the transitions between Floquet levels. For a two level system as considered here, these transitions can be divided into two types, which we treat separately:
\begin{enumerate}
    \item Transitions in which the excited state relaxes to the ground state by emission of a photon at frequency $\omega_n = \nu_k$. In this case, we have the interaction Hamiltonian 
    \begin{equation}
        V_n(t) = J_n \sum_k g_k \left(\sm b_k^\dagger e^{-i(\omega_n - \nu_k)t} + \sp b_k e^{i(\omega_n - \nu_k)t}\right).
    \end{equation}
    By using this Hamiltonian in Eq. \ref{eq:reduced_dm_iterative}, we find
    \begin{equation}
        \dot{\rho}(t)  =  -\nth \frac{\Gamma_n}{2}\left(\rho\sm\sp - 2\sp\rho\sm + \sm\sp\rho\right)
                    - (\nth + 1)\frac{\Gamma_n}{2}(\sp\sm\rho - 2\sm\rho\sp + \rho\sp\sm),
    \end{equation}
    where $\Gamma_n \equiv 2\pi g(\omega_n)^2 \rho(\omega_n)$ is a modified relaxation rate defined in terms of the coupling coefficient $g$ and density of states $\rho$, both evaluated at the shifted frequency $\omega_n$. For excited to ground processes, a total decay rate can be defined as a sum of decay rates for all harmonics which allow emission (i.e. which satisfy $\omega_n > 0$). Specifically, this is
    \begin{equation}
        \Gamma_e \equiv \sum_{\omega_n > 0} \Gamma_n
    \end{equation}
    We also note that the decay rates $\Gamma_n$ referenced here are identical to those that would be calculated by use of the Floquet Fermi Golden rule for transitions between Floquet states \cite{rudner2020floquet}.
    \item Transitions in which the ground state relaxes to the excited state by emission of a photon at $-\omega_m = \nu_k$. In the absence of time dependence, such a process cannot be energy conserving. However, harmonic shifts permit the existence of integers $m < 0 $ such that $-\omega_m = -\omega_0 - m\Omega > 0$. Under the RWA, these transitions are accounted for by Hamiltonian terms which are usually discarded as counter-rotating terms:
    \begin{equation}
        V_m(t) = J_m \sum_k g_k \left(\sm b_k e^{-i(\omega_m + \nu_k)t} + \sp b_k^\dagger e^{i(\omega_n + \nu_k)t}\right)
    \end{equation}
    Following a similar procedure, we find
    \begin{equation}
        \dot{\rho}(t)  =  - \nth \frac{\Gamma_m}{2}(\sp\sm\rho - 2\sm\rho\sp + \rho\sp\sm)
                    - (\nth + 1)\frac{\Gamma_m}{2}\left(\rho\sm\sp - 2\sp\rho\sm + \sm\sp\rho\right),
    \end{equation}
    The only difference between these Lindblad terms and the ones in Eq. B16 is that the $\nth$ and $\nth$ terms are switched. Also similarly to case (1), we can define a total rate of ground state decay
    \begin{equation}
        \Gamma_g \equiv \sum_{\omega_m < 0} \Gamma_m
    \end{equation}
\end{enumerate}

By taking the above manipulations to hold for all relevant values of $m$ and $n$, a Lindblad operator which incorporates all decays of the system can be constructed:
\begin{equation}
\begin{split}
    \dot{\rho}(t) = &-\nth \frac{\Gamma_e}{2}\left(\rho\sm\sp - 2\sp\rho\sm + \sm\sp\rho\right) \\
                    &- (\nth + 1)\frac{\Gamma_e}{2}(\sp\sm\rho - 2\sm\rho\sp + \rho\sp\sm) \\
                    &- \nth \frac{\Gamma_g}{2}(\sp\sm\rho - 2\sm\rho\sp + \rho\sp\sm) \\
                    &- (\nth + 1)\frac{\Gamma_g}{2}\left(\rho\sm\sp - 2\sp\rho\sm + \sm\sp\rho\right)
\end{split}
\end{equation}
Assuming no thermal background contribution ($\nth = 0$), the equations of motion for the diagonal density matrix take a particularly simple form which enable us to calculate the steady state populations of the driven system:
\begin{align}
    \dot{\rho}_{gg} &= -\Gamma_g \rho_{gg} + \Gamma_e \rho_{ee} \\
    \dot{\rho}_{ee} &= -\Gamma_e \rho_{ee} + \Gamma_g \rho_{gg}
\end{align}
By setting derivatives to zero, the population equation is easily solved: $\rho_{gg} = \Gamma_e/(\Gamma_e + \Gamma_g)$, $\rho_{ee} = \Gamma_{g}/(\Gamma_e + \Gamma_g)$. Thus the total inversion level referenced in the linear response coefficients is $\Braket{\sigma_z}_0 = \rho_{ee} - \rho_{gg} = (\Gamma_g - \Gamma_e)/(\Gamma_g + \Gamma_e)$.

\emph{Additional remarks on damping:} The above give an outline of how damping can be accounted for in Floquet systems, which a special emphasis on the implications for their optical response. We have focused here on a two-level model which is $\sigma_z$-modulated for simplicity. That said, the concepts explained here should apply more generally to multi-level systems which are modulated in more complex ways, such as the types of systems explored in electromagnetically induced transparency (EIT). It is also worth noting that issues of degeneracies (or quasi-degeneracies) can introduce substantial complications into the analysis of Floquet systems. Such cases should be treated carefully, with guidance from works on this topic such as \cite{hone2009statistical}. 

\section{Lorentz parametric oscillator derivations}

\subsection{Semiclassical approach}

The Lorentz oscillator is an important and commonly used model for dispersive dielectric functions, for example in the presence of resonances that result from optical phonons in polar insulators. In this section, we outline the semiclassical approach taken to derive the optical response of a parametric oscillator. To do so, we can write the equation of motion of a forced harmonic oscillator with a resonance frequency which is perturbed by some amount $f(t)$:
\begin{equation}
	\ddot{x}(t) + \Gamma\dot{x}(t) + \omega_0^2(1 + f(t))x(t) = F(t).
\end{equation}
In the absence of any parametric driving, this equation of course has the Fourier domain solution $x(\omega) = F(\omega)/(\omega_0^2 - \omega^2 - i\omega\Gamma).$

To proceed in the parametric case, we can solve for the Green's function. The time domain Green's function $K(t,t')$ is defined to satisfy:
\begin{equation}
	\left[\partial_t^2 + \Gamma\partial_t + \omega_0^2(1 + f(t))\right]K(t,t') = \delta(t-t').
\end{equation}
Rather than solving for $K(t,t')$ in the time domain and the transforming into frequency space, we will find it more useful to transform into frequency space and then consider perturbative solutions for $K(\omega, \omega')$ directly. Using the Fourier transform convention $f(\omega,\omega') = \int dt\,dt'\, e^{i\omega t} f(t,t') e^{-i\omega't'}$, we find that:
\begin{equation}
	\left[\omega_0^2 - \omega^2 - i\omega\Gamma\right]K(\omega,\omega') + \omega_0^2\int\frac{d\omega''}{2\pi} \tilde{f}(\omega - \omega'') K(\omega'', \omega') = 2\pi \delta(\omega - \omega').
\end{equation}
By solving perturbatively, we obtain the correction:
\begin{equation}
K(\omega, \omega') = \frac{2\pi\delta(\omega - \omega')}{\omega_0^2 - \omega^2 - i\omega\Gamma} - \frac{\omega_0^2 \tilde{f}(\omega - \omega')}{(\omega_0^2 - \omega^2 - i\omega\Gamma)(\omega_0^2 - \omega'^2 - i\omega'\Gamma)} + \mathcal{O}(f^3).
\end{equation}
For our purposes, this form is suitable.
\vspace{1cm}
\textit{Evaluation in the time domain for parametric oscillator: }

A particularly important case of this is the case of parametric resonance in which the driving frequency is twice the resonance frequency. To aid this, we define
\begin{equation}
	\chi_\pm^{(1)}(t,t') = -\delta f \omega_0^2 \int \frac{d\omega\,d\omega'}{(2\pi)^2}e^{-i\omega t}\frac{\pi\delta(\omega-\omega'\pm\Omega)}{(\omega_0^2 - \omega^2 - i\omega\Gamma)(\omega_0^2 - \omega'^2 - i\omega'\Gamma)}e^{i\omega't'}.
\end{equation}
We begin by completing the $\omega'$ delta function integral which sets $\omega' \to \omega\pm\Omega$. This gives
\begin{align}
	\chi_\pm^{(1)}(t,t') &= -\frac{1}{2}\delta f\omega_0^2 e^{\pm i\Omega t'}\int \frac{d\omega}{2\pi} e^{-i\omega(t-t')} \frac{1}{(\omega_0^2 - \omega^2 - i\omega\Gamma)}\frac{1}{(\omega_0^2 - (\omega\pm\Omega)^2 - i(\omega\pm\Omega)\Gamma)} \\
			&= -\frac{1}{2}\delta f\omega_0^2 e^{\pm i\Omega t'}\int d\tau\, e^{\pm i\Omega\tau} \chi^{(0)}(t-t'-\tau)\chi^{(0)}(\tau) \\
			&= -\frac{\delta f\omega_0^2 e^{\pm i \Omega t'} e^{-\frac{\Gamma}{2}(t-t')}}{2\omega_r^2}\int_0^\infty d\tau\, \theta(t-t'-\tau) e^{\pm i \Omega \tau}\sin[\omega_r(t-t'-\tau)]\sin[\omega_r\tau],
\end{align}	
where we have used $\omega_r = \frac{1}{2}\sqrt{4\omega_0^2-\Gamma^2}$, and taken advantage of convolutions, shifting frequency arguments, and the unperturbed harmonic oscillator response function $\chi^{(0)}(t-t')$. Performing the integration, and then adding the positive and negative frequency contribution as $\chi^{(1)}(t,t') = \chi_+^{(1)}(t,t') + \chi_-^{(1)}(t,t')$ gives the final result
\begin{align}
\chi^{(1)}(t,t') &= \frac{2 \delta f \omega_0^2}{\Omega  \omega_r \left(4\omega_r^2 - \Omega ^2\right)} \theta(t-t') e^{-\frac{1}{2} \Gamma  (t-t')} \cos \left(\frac{\Omega}{2}(t+t')\right) \\ &\times\left[2\omega_r \sin\left(\frac{\Omega}{2} (t-t')\right) \cos \left(\omega_r (t-t')\right)-\Omega  \cos \left(\frac{\Omega}{2}(t-t')\right) \sin \left(\omega_r (t-t')\right)\right].
\end{align}
The expression has several encouraging features. First, the expression is manifestly causal with the $\theta(t-t')$ dependence. Secondly, all terms have an explicit $(t-t')$ dependence, \emph{except for the $\cos(\Omega(t+t')/2)$ term}, which comes explicitly from the broken time translation invariance of the system. Additionally, the expression contains explicit exponential decay set by the loss parameter $\Gamma$, just as $\chi^{(0)}(t-t')$ does. 


\subsection{Quantum mechanical approach}

We now provide a quantum derivation of the Lorentz parametric oscillator susceptibility. We do this by calculating the dipole susceptibility $\alpha(t,t')$ for a quantum mechanical parametric oscillator, and showing that it matches the semiclassical approach at first order in perturbation theory. We start with a Hamiltonian of a one dimensional quantum harmonic oscillator with a resonant frequency $\omega_0$ which oscillates at frequency $\Omega$.  
\begin{equation}
    H(t) = \frac{p^2}{2m} + \frac{1}{2}m\omega_0^2\left(1 + \delta\omega\cos\Omega t\right)x^2.
\end{equation}
In the basis of the unperturbed Harmonic oscillator, we can write this in terms of creation and annihilation operators as
\begin{equation}
    H(t) = \hbar\omega_0 a^\dagger a + \frac{\hbar\omega_0}{4}\delta\epsilon\cos\Omega t \left(a+a^\dagger\right)^2.
\end{equation}
The Floquet states, to first order in $\delta\epsilon$, are given as 
\begin{equation}
    \ket{\psi_n^{(1)}(t)} = e^{-i\omega_n t}\left[(1 + S_n)\ket{n} + P_{n-1}\ket{n-2} + - P_{n+1}^*\ket{n+2} \right]
\end{equation}
where we defined $S_n(t) \equiv -i\delta\epsilon (2n+1)\eta_0(t)$ and $P_n(t) \equiv -i\delta\epsilon\sqrt{n(n+1)}\eta_1(t)$, where
\begin{equation}
    \eta_0(t) = \frac{\omega_0 \sin\Omega t}{4\Omega}, \hspace{1cm} \eta_1(t) = \frac{\omega_0}{4}\frac{\Omega\sin\Omega t - 2i\omega_0 \cos\Omega t}{\Omega^2-4\omega_0^2}
\end{equation}
Using this, we can write the unitary time evolution order to first order as 
\begin{equation}
    U(t) = \sum_n e^{-i\omega_n t}\left[(1 + S_n)\ket{n}\bra{n} + P_{n-1}\ket{n-2}\bra{n} - P_{n+1}^*\ket{n+2}\bra{n}\right].
\end{equation}
Then we need the dipole moment $d(a+a^\dagger)$ in the interaction picture (i.e. transformed by $U(t)$). Doing this, and discarding terms greater than first order in $\delta\epsilon$, we find
\begin{equation}
    U^\dagger(t) (a+a^\dagger) U(t) = \sum_{n}\sum_{k=\pm 1, \pm 3} e^{-i(\omega_n-\omega_{n+k})t} A_k \ket{n+k}\bra{n}
\end{equation}
where 
\begin{align}
    A_1 &= \left(1 + S_{n+1}^*\right)\left[\sqrt{n+1}\left(1+S_n\right) - \sqrt{n+2} P_{n+1}^*\right] + P_n^* \left[\sqrt{n}\left(1+S_n\right) + \sqrt{n-1}P_{n-1}\right] \\
    A_{-1} &= \left(1 + S_{n-1}^*\right)\left[\sqrt{n}\left(1+S_n\right) + \sqrt{n-1} P_{n-1}\right] - P_n \left[\sqrt{n+1}\left(1+S_n\right) - \sqrt{n+2}P_{n+1}^*\right] \\
    A_3 &= -\sqrt{n+3}P_{n+1}^* + P_{n+2}^*\left[\sqrt{n+1}(1+S_n) - \sqrt{n+2}P_{n+1}^*\right] \\
    A_{-3} &= \sqrt{n-2}P_{n-1} - P_{n-2}\left[\sqrt{n}(1+S_n) + \sqrt{n-1}P_{n-1}\right]
\end{align}
Then we find that
\begin{equation}
    \Braket{0|d_I(t)d_I(t')|0} = d^2 e^{-i\omega_0(t-t')}\left(1 + S_0^*(t) + S_1(t) - \sqrt{2} P_1(t)\right)\left(1 + S_0(t') + S_1^*(t') - \sqrt{2} P_1^*(t')\right)
\end{equation}
Several of these terms are second order in $\delta\epsilon$ and can be discarded. Doing this, the commutator for linear response is
\begin{equation}
     \Braket{0|\left[d_I(t,)d_I(t')\right]|0} = \left(e^{-i\omega_0(t-t')} - \text{c.c}\right) - 2i\delta\epsilon d^2 e^{-i\omega_0(t-t')}\left[\eta_0(t) - \eta_0(t') - \left(\eta_1(t) - \eta_1^*(t')\right)\right] - \text{c.c}
\end{equation}
Then the polarizability to first order can be expressed as 
\begin{equation}
    \alpha(t,t') = \alpha^{(0)}(t-t') + \alpha^{(1)}(t,t') + \mathcal{O}(\delta\epsilon^2)
\end{equation}
where
\begin{equation}
    \alpha^{(0)}(t-t') = \frac{2d^2}{\hbar}\theta(t-t')\sin\left(\omega_0(t-t')\right)
\end{equation}
and
\begin{equation}
\begin{split}
    \alpha^{(1)}(t, t') &= -\frac{4\delta\omega d^2 \omega_0^2}{\hbar\Omega \left(\Omega^2 - 4\omega_0^2\right)}\theta(t-t')\cos\left(\frac{\Omega}{2}(t+t')\right) \\
    &\times \left[2\omega_0\sin\left(\frac{\Omega}{2}(t-t')\right) \cos\left(\omega_0(t-t')\right) - \Omega \cos\left(\frac{\Omega}{2}(t-t')\right) \sin\left(\Omega_0(t-t')\right)  \right] 
\end{split}
\end{equation}
In the limit of no dissipation, this is equivalent to the Lorentz parametric oscillator model that was derived semiclassically. 

\section{Electrodynamics with time-varying materials}

\subsection{Maxwell's Equations in Time Dependent Materials}

In this section, we will formulate Maxwell's equations in media which are time-periodic and dispersive. By making the assumption of a spatially homogeneous medium, we will be able to cast Maxwell's equations into a Floquet eigenvalue problem for the wavevectors and quasifrequencies which can propagate. By solving this problem numerically, we can compute band structures. In the absence of external sources, the Maxwell equation for the electric field can be written as
\begin{equation}
	\curl\curl\Ev(\rv, \omega) - \frac{\omega^2}{c^2}\int \frac{d\omega'}{2\pi} \varepsilon(\rv, \omega,\omega') \Ev(\rv, \omega') = 0.
\end{equation}
If we write epsilon as time-independent background $\varepsilon_{\text{bg}}(\omega)$ plus a perturbation $\Delta\chi(\omega,\omega')$ which results from a time dependence, we can write $\varepsilon(\rv, \omega, \omega') = \varepsilon_{\text{bg}}(\rv, \omega)[2\pi\delta(\omega-\omega')] + \Delta\chi(\rv, \omega,\omega')$. Substituting this into the Maxwell equation, we obtain the form
\begin{equation}
    \curl\curl\Ev(\rv, \omega) - \frac{\omega^2}{c^2}\varepsilon_{\text{bg}}(\rv, \omega)\Ev(\rv, \omega)
    = \frac{\omega^2}{c^2}\int \frac{d\omega'}{2\pi} \Delta\chi(\rv, \omega,\omega') \Ev(\rv, \omega').
\end{equation}
If the driving is periodic, and the medium is assumed be uniform in space, then the response function can be cast into the form
\begin{equation}
	\Delta\chi(\omega,\omega') = \sum_k \Delta\chi_k(\omega)2\pi\delta(\omega-\omega'-k\Omega_0).
\end{equation}
Substituting this into the Maxwell equation gives
\begin{equation}
	\curl\curl\Ev(\omega) - \frac{\omega^2}{c^2}\varepsilon_{\text{bg}}(\omega)\Ev(\omega) = \frac{\omega^2}{c^2}\sum_k \Delta\chi_k(\omega)\Ev(\omega-k\Omega_0).
\end{equation}
Assuming a bulk medium which can be spatially decomposed into plane waves, and invoking the Bloch-Floquet requirement for the time-dependent portion of the mode functions, we can write
\begin{equation}
	\Ev(\rv, t) = e^{i \kv\cdot\rv}\sum_n u_{\Omega n} e^{-i(\Omega+n\Omega_0)t} \implies \Ev(\rv,\omega) = e^{i \kv\cdot\rv} 2\pi\sum_n u_{\Omega n}\delta(\omega-\Omega-n\Omega_0).
\end{equation}
For convenience, we will define $\Omega_n \equiv \Omega + n\Omega_0$. Substituting this into the Maxwell equation, integrating both sides in $\omega$ to isolate frequency components, and relabeling sums gives the final central equation result
\begin{equation}
	\left[k^2 - \frac{\Omega_n^2}{c^2}\varepsilon_{\text{bg}}(\Omega_n)\right]u_n = \frac{\Omega_n^2}{c^2}\sum_m \Delta\chi_m(\Omega_n)u_{n-m}.
\end{equation}
Rewriting this as an eigenvalue problem for $k$ in terms of $\Omega$, we have
\begin{equation}
	\frac{\Omega_n^2}{c^2}\varepsilon_{\text{bg}}(\Omega_n)u_n + \frac{\Omega_n^2}{c^2}\sum_m \Delta\chi_m(\Omega_n)u_{n-m} = k_\Omega^2 u_n,
	\label{eq:maxwell_floquet_eigenvalue_appendix}
\end{equation}
which casts the dispersion as a relatively simple central equation eigenvalue problem. Eq. \ref{eq:maxwell_floquet_eigenvalue} can be easily implemented as a matrix eigenvalue problem, yielding the wavevector $k_\Omega$ and Floquet mode amplitudes $u_{\Omega n}$ at each quasifrequency $\Omega \in [-\Omega_0/2, \Omega_0/2)$.

\subsection{Reflection and Transmission}

Once the Maxwell Floquet solutions have been obtained, as described in the previous section, they can be used to solve classical optics problems. In this section, we show the illustrative example of a reflection-transmission problem of a monochromatic field incident on a slab of material characterized by a time-dependent linear response function.

We write the Floquet modes as
\begin{equation}
	u_{b,\tilde{\Omega}}(t) = \sum_m u_{b,\tilde{\Omega}}^{(m)} e^{-i\Omega_0 m t},
\end{equation}
where $b$ is a band index, and $\sum_m$ is a sum over harmonics of the driving frequency $\Omega_0$. Next we can consider an interface between two materials (1) and (2), where (1) for now is air, and (2) is a material described with this kind of formalism, and we assume we have solved for the bands. Furthermore, we assume that the electric field is polarized in the plane of the interface (s-polarized). 

In this analysis, it will be helpful to express the incident frequency as $\omega_0 = \tilde{\omega}_0 + s\Omega_0$, where $\tilde{\omega}_0$ is confined to lie in the first Brillouin zone $-\frac{\Omega_0}{2} < \tilde{\omega}_0 < \frac{\Omega_0}{2}$. In this case, $s$ is easily interpreted as a number of harmonics by which the true incoming frequency is offset from the BZ frequency that is seen in the time-dependent medium. Thus we can write the incident field as 
\begin{equation}
	E^{\text{inc}}(x,t) = e^{ik_0 x}e^{-i\omega_0 t},
\end{equation}
where $\omega_0$ is the frequency of the incident plane wave, and $k_0 = k(\omega_0) = \omega_0/c$ is the vacuum dispersion. For now, we assume that the reflected field can have any frequency, so we can write very generally
\begin{equation}
	E^{(\text{r})}(x,t) = \sum_{\omega > 0} r(\omega) e^{-ik(\omega)x} e^{-i\omega t}.
\end{equation}
Finally, the transmitted field is
\begin{align}
	E^{(\text{t})}(x,t) &= \sum_b t_b e^{iq_b(\tilde{\omega}_0 x} e^{-i\tilde{\omega}_0 t} u_{b,\tilde{\omega}_0}(t) \\
		&= \sum_{b,m} t_b u_{b,\tilde{\omega}_0}^{(m)} e^{iq_b(\tilde{\omega}_0) x} e^{-i(\tilde{\omega}_0 + m\Omega_0) t}.
\end{align}

The boundary conditions at the interface are given by $E^{\text{inc}}(0,t) + E^{\text{r}}(x,t) = E^{\text{t}}(0,t)$ and
$\partial_x E^{\text{inc}}(x,t)|_{x=0} + \partial_x E^{\text{r}}(x,t)|_{x=0} = \partial_x E^{\text{t}}(x,t)|_{x=0}$. We can write a matrix equation down for the boundary conditions. Now we have boundary conditions at $x=\pm L/2$. The first two rows are for continuity of the field at $\pm L/2$ respectively. The last two rows are for continuity of the derivative. Doing this, we find the matrix equations
\begin{equation}
\begin{pmatrix}
-I*e^{ik_0(-L/2)} & \mathbf{u}_b e^{ik_b(\tilde{\omega}_0)(-L/2)} & \mathbf{u}_b e^{-ik_b(\tilde{\omega}_0)(-L/2)} & \mathbf{0} \\
\mathbf{0} & \mathbf{u}_b e^{ik_b(\tilde{\omega}_0)(L/2)} & \mathbf{u}_b e^{-ik_b(\tilde{\omega}_0)(L/2)} & -I*e^{ik_0(L/2)} \\
(\Omega_{m}/c)e^{i(\Omega_m/c)(-L/2)} & k_b(\tilde{\omega_0})\mathbf{u}_be^{ik_b(\tilde{\omega}_0)(-L/2)} & -k_b(\tilde{\omega_0})\mathbf{u}_be^{-ik_b(\tilde{\omega}_0)(-L/2)} & \mathbf{0} \\
\mathbf{0} & k_b(\tilde{\omega_0})\mathbf{u}_be^{ik_b(\tilde{\omega}_0)(L/2)} & -k_b(\tilde{\omega_0})\mathbf{u}_be^{-ik_b(\tilde{\omega}_0)(L/2)} & (\Omega_{m}/c)e^{i(\Omega_m/c)(L/2)}
\end{pmatrix}
\begin{pmatrix}
    \mathbf{r} \\ \mathbf{a} \\ \mathbf{b} \\ \mathbf{t}
\end{pmatrix} = 
\begin{pmatrix}
    \mathbf{v}_1 \\ \mathbf{v}_2 \\ \mathbf{v}_3 \\ \mathbf{v}_4
\end{pmatrix}
\end{equation}

Finally, we have $\mathbf{v}_2 = \mathbf{v}_4 = 0$. Then we have $(\mathbf{v}_1)_s = e^{ik_0(-L/2)}$ and $(\mathbf{v}_3)_s = (\Omega_s/c)e^{i(\Omega_s/c)(-L/2)}$, where the $s$ index refers to the index that matches $s$ (so actually $M+1+s$). By performing matrix inversion, one can find the reflected, transmitted, and internal fields for the scattering problem which is described in Fig. 4 of the main text.

\end{document}


\rmfamily

\title{Supplementary Information: \\
Dispersion in photonic time crystals}
\author{Jamison Sloan$^{1\dagger}$, Nicholas Rivera$^{2}$, John D. Joannopoulos$^{2}$, and Marin Solja\v{c}i\'{c}$^{2}$}

\affiliation{$^{1}$ Department of Electrical Engineering and Computer Science, Massachusetts Institute of Technology, Cambridge, MA 02139, USA \\
$^{2}$ Department of Physics, Massachusetts Institute of Technology, Cambridge, MA 02139, USA
$\dagger$ Corresponding author e-mail: jamison@mit.edu}

\noindent

\maketitle
\tableofcontents
\newpage 

\section{General Properties of Linear response}

\subsection{Kubo formula for Floquet systems}

Here, we will derive an expression for the two-frequency atomic polarizability $\alpha(\omega, \omega')$ of a generic single-electron system which is varied in time. We assume that the unperturbed system is described by a Hamiltonian $H_0(t)$, with the corresponding time evolution operator $U_0(t)$ which satisfies $i\hbar\partial_t U_0(t) = H_0(t)U_0(t)$  We can send an electric field probe which couples to the dipole operator $d$ of the unperturbed system as $V(t) = -d E(t)$. Here, we have written our expressions in terms of a single coordinate, but all that follows can be easily generalized to include vector directions for fields and dipole moments. Then, by linear response theory, the change in the dipole moment is given as 
\begin{equation}
    \delta\Braket{d(t)} = \int_{-\infty}^\infty dt'\,\alpha(t, t') E(t'),
\end{equation}
where the Kubo formula for the two-time polarizability is given as 
\begin{equation}
    \alpha(t,t') = \frac{i}{\hbar}\theta(t-t')\Braket{\psi_0|[d_I(t), d_I(t')]|\psi_0}.
\end{equation}
Here, $d_I(t) = U_0^\dagger(t) d U_0(t)$ is the dipole moment operator in the interaction picture, and $\ket{\psi_0}$ is the state of the system before the probe is applied. In the frequency domain, we can write
\begin{equation}
    \delta\Braket{d(\omega)} = \int_{-\infty}^\infty \frac{d\omega'}{2\pi}\alpha(\omega, \omega') E(\omega'),
\end{equation}
where $\alpha(\omega, \omega') \equiv \int dt\,dt'\, e^{i(\omega t - \omega't')} \alpha(t,t')$. 

In general, $\alpha(\omega, \omega')$ can be a function of the two continuous frequencies $\omega, \omega'$. However, this work will focus primarily on systems where the system is time-modulated periodically. This will allow us to make key simplifications. Namely, Floquet theory will be used to decompose the problem into harmonics. In this case, we assume that the time dependent Hamiltonian $H_0(t)$ has a period $T$, so that $H_0(t + T) = H_0(t)$. The frequency associated with this period is $\Omega_0 \equiv 2\pi/T$. In this case, solutions to the time-dependent Schrodinger equation $i\hbar\partial_t\ket{\psi_\alpha(t)} = H_0(t)\ket{\psi_\alpha(t)}$ can be written in terms of Floquet states
\begin{equation}
    \ket{\psi_\alpha(t)} = e^{-i\epsilon_\alpha t} \ket{\phi_\alpha(t)}
\end{equation}
where $\epsilon_\alpha$ is the Floquet quasi-energy which lies in the Brillouin zone, and $\ket{\phi_\alpha(t)}$ is a periodic function (known as a Floquet mode) which can be decomposed in terms of harmonics as $\ket{\phi_\alpha(t)} = \sum_n e^{in\Omega_0 t}\ket{\phi_\alpha^n}$

By assuming periodicity, we can write a form for $\alpha(t,t')$ in terms of the Floquet modes. By inserting a complete set of Floquet states $\ket{\psi_\alpha(0)}$, we can write
\begin{align}
    \Braket{\psi_0|d_I(t)d_I(t')|\psi_0} &= \sum_\alpha \Braket{\psi_0|U_0^\dagger(t) d U_0(t) |\psi_\alpha} \Braket{\psi_\alpha|U_0^\dagger(t') d U_0(t') |\psi_0} \\
    &= \sum_\alpha \Braket{\psi_0(t)|d|\psi_\alpha(t)}\Braket{\psi_\alpha(t')|d|\psi_0(t')}
\end{align}

Substituting these states into the above expression for dipole expectation value, we find
\begin{equation}
    \Braket{\psi_0|d_I(t)d_I(t')|\psi_0} = \sum_\alpha \sum_{k,l,m,n} e^{-i(\epsilon_\alpha - \epsilon_0)t + i(n-k)\Omega t} e^{i(\epsilon_\alpha-\epsilon_0)t' - i(m-l)\Omega t'} \Braket{\phi_0^k|d|\phi_\alpha^n}\Braket{\phi_\alpha^m|d|\phi_0^l}
\end{equation}
Then we can use $\theta(t-t') = i\int\frac{d\omega''}{2\pi}\frac{e^{-i\omega''(t-t')}}{\omega'' + i\eta}$ to write 
\begin{equation}
    \begin{split}
        \frac{i}{\hbar} \int dt\,dt'\,e^{i\omega t - i\omega't'}\theta(t-t')& \Braket{\psi_0|d_I(t)d_I(t')|\psi_0} 
        = -\frac{1}{\hbar} \int\frac{d\omega''}{2\pi}\frac{1}{\omega'' + i\eta}\int dt\,dt'\,e^{i\omega t - i\omega't'} e^{-i\omega''(t-t')} \\
        &\times\left(\sum_\alpha \sum_{k,l,m,n} e^{-i(\epsilon_\alpha - \epsilon_0)t + i(n-k)\Omega t} e^{i(\epsilon_\alpha-\epsilon_0)t' - i(m-l)\Omega t'} \Braket{\phi_0^k|d|\phi_\alpha^n}\Braket{\phi_\alpha^m|d|\phi_0^l}\right)
    \end{split}
\end{equation}
Turning the exponentials into delta functions, completing the $\omega''$ integral trivially, and doing all of the same for the other term, we find
\begin{equation}
\begin{split}
    \alpha(\omega, \omega') = -\frac{2\pi}{\hbar}\sum_\alpha \sum_{k,l,m,n} &\left(\frac{\Braket{\phi_0^k|d|\phi_\alpha^n}\Braket{\phi_\alpha^m|d|\phi_0^l}}{\omega + (n-k)\Omega - (\epsilon_\alpha - \epsilon_0) + i\eta} - \frac{\Braket{\phi_0^k|d|\phi_\alpha^n}\Braket{\phi_\alpha^m|d|\phi_0^l}}{\omega - (m-l)\Omega + (\epsilon_\alpha - \epsilon_0) + i\eta}\right) \\
    &\hspace{2cm}\times \delta\left[\omega - \omega' - (m+k-n-l)\Omega\right]
\end{split}   
\label{eq:floquet_kubo}
\end{equation}
We immediately note that $\alpha(\omega, \omega')$ takes the form of a sum over delta functions of the form $\delta(\omega-\omega'-k\Omega_0)$, where $k$ is an integer. In other words, an incident field at frequency $\omega'$ can induce a dipole moment at frequencies $\omega' +k\Omega_0$. This property emerges solely from assuming that the system is periodic. As such, for periodic systems, we may replace $\alpha(\omega, \omega')$ with a series of single-argument functions $\alpha_k(\omega)$ defined by
\begin{equation}
    \alpha(\omega, \omega') = \sum_k \alpha_k(\omega) 2\pi\delta(\omega-\omega'-k\Omega_0).
\end{equation}

\subsection{General properties of $\chi(\omega,\omega')$}

We assume that some source $F(t)$ and some observable $x(t)$ are related by a linear response function $\chi(t,t')$ as
\begin{equation}
	x(t) = \int \chi(t,t') F(t')\,d t',
\end{equation}
so that in frequency space we have
\begin{equation}
	x(\omega) = \frac{1}{2\pi} \int \chi(\omega,\omega')F(\omega')\, d \omega'.
\end{equation}

Here we discuss the general properties of two-frequency susceptibility functions using as few assumptions as possible. We assume that $\chi(\omega,\omega')$ is expressed as the Fourier transform of some \emph{real-valued} two-time correlation function $\chi(t,t')$. We may also assume that $\chi(t,t') \propto \theta(t-t')$ to preserve causality. The Fourier transform is expressed as 
\begin{equation}
	\chi(\omega, \omega') = \int dt\,dt'\, e^{i(\omega t + \omega't')} \chi(t,t').
	\label{eq:two_freq_ft}
\end{equation}
Now we list some basic properties that follow from these assumptions.
\begin{enumerate}
\item Double negation. From the form of \ref{eq:two_freq_ft}, we see that negating \emph{both} frequency arguments changes the sign of the phase in the complex exponential. Thus $\chi(-\omega,-\omega') = \chi^*(\omega,\omega')$.
\item Single negation. Similarlly, negating a single frequency argument is equivalent to negating the other frequency argument and then conjugating. This is $\chi(-\omega, \omega') = \chi^*(\omega, -\omega')$.
\item Real an imaginary parts. Taking the Real and Imaginary parts of the previous claim gives $\text{Re}\chi(-\omega,\omega') = \text{Re}\chi(\omega,-\omega')$, and $\text{Im}\chi(-\omega,\omega') = -\text{Im}\chi(\omega,-\omega')$
\end{enumerate}

\subsection{Kramers Kronig relations in time-dependent systems}

In this section, we derive Kramers Kronig (K.K.) relations for linear response functions $\chi(\omega, \omega')$ which are not time-translation invariant. The conventional proof of Kramer's Kronig relations uses complex analysis to show that optical passivity of a linear response function $\chi(\omega)$ implies to the relation
\begin{equation}
    \chi(\omega) = \frac{i}{\pi}\mathcal{P}\int d\omega''\frac{\chi(\omega'')}{\omega - \omega''},
    \label{eq:conventional_kk}
\end{equation}
where $\mathcal{P}$ denotes the principle value of the integral. Then, splitting this equation into complex components yields a direct relationship between the real and imaginary parts of $\chi(\omega)$. In a passive system, $\Im \chi(\omega)$ encodes the dissipation of the material. A basic consideration of energy conservation requires that $\Im \chi(\omega) \geq 0$ for all $\omega > 0$ to ensure that inputs are attenuated over time, rather than amplified. While the conventional K.K. relation is usually explained in terms of optical passivity (and complex poles in the upper half plane), Eq. \ref{eq:conventional_kk} can also be derived as an immediate consequence of \emph{causality}: the fact that a system can only respond to an impulse after its application.

While linear response functions in time-varying systems are not necessarily passive, they must respect causality. We will use this constraint to derive K.K. relations for linear response functions in time-dependent systems. For a time-dependent response function $\chi(t,t')$, the causality constraint can be expressed in terms of the heaviside function $\theta(t)$ as
\begin{equation}
    \chi(t, t') = \theta(t -t')\chi(t, t').
    \label{eq:two_time_causaility}
\end{equation}
To derive the K.K. relation, we take the two-time Fourier transform of both sides, using the convention $\chi(\omega, \omega') = \int dt\,dt'\,e^{i(\omega t - \omega't')}\chi(t,t')$. This allows us to write
\begin{equation}
    \chi(\omega, \omega') = \int dt\,dt'\, e^{i(\omega t - \omega't')} \theta(t-t')\chi(t,t').
    \label{eq:two_time_causality_fourier}
\end{equation}
The right hand side can be evaluated using the convolution theorem. To perform this, we note that the double Fourier transform of the heaviside function is given by
\begin{equation}
    \theta(\omega, \omega') = \int dt\,dt'\,e^{i(\omega t - \omega't')}\theta(t-t') = 2\pi\delta(\omega-\omega')\left[2\pi\delta(\omega+\omega') + \frac{2i}{\omega+\omega'}\right].
\end{equation}
Then the convolution integral gives
\begin{equation}
\begin{split}
    \int dt\,dt'\, e^{i(\omega t - \omega't')} \theta(t-t')\chi(t,t') &= \int \frac{d\nu\,d\nu'}{(2\pi)^2} \chi(\omega - \nu, \omega' - \nu')\theta(\nu, \nu') \\
    &= \frac{1}{2\pi}\int d\nu \chi(\omega - \nu, \omega' - \nu)\left[2\pi(\nu + \nu') + \frac{i}{\nu}\right] \\
    &= \frac{1}{2}\chi(\omega, \omega') + \frac{i}{2\pi}\int d\nu\, \frac{\chi(\omega - \nu, \omega' - \nu)}{\nu}
\end{split}
\end{equation}
Plugging this back into Eq. \ref{eq:two_time_causality_fourier}, we can solve for $\chi(\omega, \omega')$ to give the K.K. relation
\begin{equation}
    \chi(\omega, \omega') = \frac{i}{\pi}\mathcal{P}\int d\omega'' \frac{\chi(\omega - \omega'', \omega' - \omega'')}{\omega''}.
    \label{eq:new_kk}
\end{equation}
To see how this general relation relates to the usual time-translation-invariant case, we take $\chi(\omega, \omega') = 2\pi\delta(\omega - \omega')\chi(\omega)$. Substituting this form into the new relation gives $\chi(\omega) = \frac{i}{\pi}\int d\omega'' \frac{\chi(\omega - \omega'')}{\omega''}$, which after changing the integration variable, matches the usual form noted in Eq. \ref{eq:conventional_kk}. The new K.K. given in Eq. \ref{eq:new_kk} importantly indicates that even without passivity, causality still requires a strict relation between the real and imaginary parts of $\chi(\omega, \omega'$. Specifically, these are:
\begin{align}
    \Re \chi(\omega, \omega') &= -\frac{1}{\pi} \mathcal{P}\int d\omega'' \frac{\Im\chi(\omega - \omega'', \omega' - \omega'')}{\omega''} \\
    \Im \chi(\omega, \omega') &= \frac{1}{\pi} \mathcal{P}\int d\omega'' \frac{\Re\chi(\omega - \omega'', \omega' - \omega'')}{\omega''}.
\end{align}

While the K.K. relation Eq. \ref{eq:new_kk} is valid for any linear time-varying system, we can make simplifications in the case where the time-varying system is periodic with frequency $\Omega_0$. In this case, we have seen that the response function necessarily takes the form of a sum over harmonics $\chi(\omega, \omega') = \sum_k \chi_k(\omega) 2\pi\delta(\omega-\omega'-k\Omega_0)$. Plugging this assumption into Eq. \ref{eq:new_kk} shows that the harmonics behave independently from one another. Specifically, each harmonic component $\chi_k(\omega)$ actually satisfies the time-invariant K.K. relation 
\begin{equation}
    \chi_k(\omega) = \frac{i}{\pi}\mathcal{P}\int d\omega''\frac{\chi_k(\omega'')}{\omega - \omega''}.
\end{equation}

\section{Example models}

\subsection{2 Level system}

The atomic polarizability is time domain is given via the Kubo formalism as 
\begin{equation}
    \alpha(t,t') = \frac{i}{\hbar}\theta(t-t')\braket{\psi_0|[d_I(t), d_I(t')]|\psi_0},
    \label{eq:kubo_time_domain}
\end{equation}
where $d_I(t)$ is the dipole moment operator in the interaction picture of the time dependent Hamiltonian in absence of the probe field. We assume that the Hamiltonian of the driven two-level system takes the form
\begin{equation}
    H_0(t) = \frac{\sigma_z}{2}\left(\omega_0 + \delta\omega\cos \Omega t\right).
    \label{eq:2LS_hamiltonian}
\end{equation}
This Hamiltonian commutes with itself at all times, which means that the unitary evolution operator can be evaluated directly without any concerns of time ordering as 
\begin{equation}
    U_0(t) = \exp\left[-i\int^t dt'\,H_0(t')\right] = e^{-i\omega_0 \sigma_z t/2} e^{-i(\delta\omega/2\Omega)\sigma_z \sin\Omega t}
\end{equation}
This special case enables us to compute the polarizability $\alpha(\omega, \omega')$ using two complementary approaches. In the first approach, we use Eq. \ref{eq:floquet_kubo} along with the Floquet states for the Hamiltonian to obtain $\alpha(\omega, \omega')$ directly. In the second approach, we calculate the interaction picture dipole operator $d_I(t)$ and evaluate $\alpha(t,t')$ directly from the commutator form of the Kubo formula (Eq. \ref{eq:kubo_time_domain}). 

\subsubsection{Floquet state approach}
We note that the second can be expanded as a Floquet series using the Jacobi-Anger expansion $e^{iz\sin\theta} = \sum_n J_n(z) e^{in\theta}$. This means that we can write expressions for the time-dependent Floquet states
\begin{align}
    \ket{g(t)} &= e^{-i\omega_0 t/2} \sum_m e^{im\Omega t} J_m\left(-\frac{\delta\omega}{2\Omega}\right)\ket{g} \\
    \ket{e(t)} &= e^{i\omega_0 t/2} \sum_m e^{im\Omega t} J_m\left(\frac{\delta\omega}{2\Omega}\right)\ket{e}
\end{align}
The states clearly evolve in the form of Eq. (). We see that the two Floquet states consist of the original eigenstates $\ket{g}$ and $\ket{e}$ with a time-dependent phase attached. This is a feature of this particularly simple example, as the Hamiltonian (Eq. \ref{eq:2LS_hamiltonian}) has only a $\sigma_z$ term, so the states necessarily evolve independently. Thus in this case, the Floquet levels can be written as $\ket{g_m} = J_m(-\delta\omega/2\Omega)\ket{g}$ and $\ket{e_m} = J_m(\delta\omega/2\Omega)\ket{e}$. 

\subsubsection{Direct approach}
We can now proceed quite directly with the evaluation of the dipole moment operator $d\sigma_x$ in the interaction picture. Noting that 
\begin{equation}
    e^{-i\sigma_z f(t)} \sigma_x e^{i\sigma_z f(t)} = \sigma_x \cos 2f(t) + \sigma_y \sin 2f(t)
\end{equation}
we find that $d_I(t) = d\left(\sigma_x \cos 2f(t) + \sigma_y \sin 2f(t)\right)$, and consequently
\begin{equation}
    [d_I(t), d_I(t')] = -2i d^2 \sigma_z \left(\cos 2f(t) \sin 2f(t') - \sin 2f(t)\cos 2f(t')\right) = -2id^2 \sigma_z \sin\left[2\left(f(t) - f(t')\right)\right].
\end{equation}
Using $f(t) = \omega_0 t/2 + \delta\omega/(2\Omega)\sin\Omega t$, we can write 
\begin{equation}
    \left[d_I(t), d_I(t')\right] = -2id^2\sigma_z \sin\left[\omega_0(t-t') + \frac{\delta\omega}{\Omega}\left(\sin\Omega t - \sin\Omega t'\right)\right]
\end{equation}
To proceed we note that the sine can be written in terms of complex exponentials, and then expanded in terms of Bessel functions as 
\begin{equation}
    \sin\left[\omega_0(t-t') + \frac{\delta\omega}{\Omega}\left(\sin\Omega t - \sin\Omega t'\right)\right] = \frac{1}{2i}\left[e^{i\omega_0(t-t')}\sum_{m,n} J_m\left(\frac{\delta\omega}{\Omega}\right)J_n\left(\frac{\delta\omega}{\Omega}\right)e^{in\Omega t}e^{-im\Omega t'} - \text{c.c} \right]
\end{equation}

In frequency space we have
\begin{equation}
    \alpha(\omega, \omega') = \int dt\,dt'\,e^{i\omega t} \alpha(t,t') e^{-i\omega't'}
\end{equation}
Performing this directly, we find (may be a 2 or minus missing)
\begin{equation}
    \alpha(\omega, \omega') \propto \frac{2\pi d^2}{\hbar}\sum_{m,n=-\infty}^\infty J_m\left(\frac{\delta\omega}{\Omega}\right)J_n\left(\frac{\delta\omega}{\Omega}\right) \left[\frac{\delta(\omega-\omega'+(m-n)\Omega)}{\omega-\omega_0-n\Omega+i\eta} - \frac{\delta(\omega-\omega'-(m-n)\Omega)}{\omega+\omega_0+n\Omega+i\eta}\right]
\end{equation}
When the perturbation $\delta\omega \to 0$, we recover
\begin{equation}
    \alpha(\omega, \omega') = -\frac{d^2}{\hbar} 2\pi\delta(\omega-\omega')\left[\frac{1}{\omega-\omega_0+i\eta} - \frac{1}{\omega+\omega_0+i\eta}\right]
\end{equation}
as required. For a particular $\alpha_k(\omega)$ we can write
\begin{equation}
    \alpha_k(\omega) = -\frac{d^2}{\hbar}\sum_n \left[\frac{J_n(\delta\omega/\Omega)J_{n-k}(\delta\omega/\Omega)}{\omega - \omega_0 - (n-k)\Omega + i\eta} - \frac{J_n(\delta\omega/\Omega)J_{n+k}(\delta\omega/\Omega)}{\omega + \omega_0 + (n+k)\Omega + i\eta}\right]
\end{equation}

\subsection{Lorentz parametric oscillator}

\subsubsection{Semiclassical approach}

The Lorentz oscillator is a commonly used model for dispersive dielectric functions. In this model, one considers a simple harmonic oscillator with damping
\begin{equation}
	\ddot{x}(t) + \Gamma\dot{x}(t) + \omega_0^2x(t) = F(t) \implies x(\omega) = \frac{F(\omega)}{\omega_0^2 - \omega^2 - i\omega\Gamma}.
\end{equation}
In the context of a bulk set of oscillators one typically writes
\begin{equation}
	\epsilon(\omega) = 1 + \frac{\omega_p^2}{\omega_0^2 - \omega^2 - i\omega\Gamma},
\end{equation}
where $\omega_p$ is the plasma frequency which sets the oscillator strength. Motivated by the Lorentz oscillator model, we instead consider the equation 
\begin{equation}
	\ddot{x}(t) + \Gamma\dot{x}(t) + \omega_0^2(1 + f(t))x(t) = F(t),
\end{equation}
where $f(t)$ is now an arbitary function of time. We can then think about the Green's function of this equation, which we label $K(t,t')$, by finding solutions to 
\begin{equation}
	\left[\partial_t^2 + \Gamma\partial_t + \omega_0^2(1 + f(t))\right]K(t,t') = \delta(t-t').
\end{equation}
An arguably more useful object is this Green's function fourier transformed in both time arguments to become instead a function of two frequencies $K(\omega, \omega')$. Rather than solving for $K(t,t')$ in the time domain and the transforming into frequency space, we will find it more useful to transform into frequency space and then consider perturbative solutions for $K(\omega, \omega')$ directly. Using the convention $f(\omega,\omega') = \int dt\,dt'\, e^{i\omega t} f(t,t') e^{-i\omega't'}$, the equation becomes
\begin{equation}
	\left[\omega_0^2 - \omega^2 - i\omega\Gamma\right]K(\omega,\omega') + \omega_0^2\int\frac{d\omega''}{2\pi} \tilde{f}(\omega - \omega'') K(\omega'', \omega') = 2\pi \delta(\omega - \omega').
\end{equation}
Then, solving perturbatively, we obtain
\begin{equation}
K(\omega, \omega') = \frac{2\pi\delta(\omega - \omega')}{\omega_0^2 - \omega^2 - i\omega\Gamma} - \frac{\omega_0^2 \tilde{f}(\omega - \omega')}{(\omega_0^2 - \omega^2 - i\omega\Gamma)(\omega_0^2 - \omega'^2 - i\omega'\Gamma)} + \mathcal{O}(f^3).
\end{equation}
Then the Green's function in time domain is obtained by two Fourier transforms as 
\begin{equation}
	K(t,t') = \int \frac{d\omega}{2\pi}\frac{d\omega'}{2\pi} e^{-i\omega t} K(\omega,\omega') e^{i\omega't'}.
\end{equation}
In general, this may be very difficult. However, we will study at first order in perturbation theory a case of particular interest: the parametric oscillator. For a parametric oscillator, the resonant frequency is sinusoidally modulated in time by a frequency $\Omega$. In order to aid calculations, and maintain connection to more ``physical'' scenarios, we will use an exponential regulator so that the parametric drive vanishes at $t=\pm\infty$. With these considerations, we take $f(t) = \delta f e^{-\eta|t|}\cos(\Omega t)$. In frequency space, the regulator causes delta functions to turn into Lorentzians. With our conventions, the corresponding transform is
\begin{equation}
	f_\eta(\omega) = \frac{\eta}{(\omega-\Omega)^2 + \eta^2} + \frac{\eta}{(\omega+\Omega)^2 + \eta^2},
\end{equation} 
so that $\lim_{\eta\to 0} f_\eta(\omega) = \pi\delta(\omega-\Omega) + \pi\delta(\omega+\Omega)$.

It turns out we can explicitly calculate this without the regulator, just using the delta functions. We define
\begin{equation}
	\chi_\pm^{(1)}(t,t') = -\delta f \omega_0^2 \int \frac{d\omega\,d\omega'}{(2\pi)^2}e^{-i\omega t}\frac{\pi\delta(\omega-\omega'\pm\Omega)}{(\omega_0^2 - \omega^2 - i\omega\Gamma)(\omega_0^2 - \omega'^2 - i\omega'\Gamma)}e^{i\omega't'}.
\end{equation}
We begin by completing the $\omega'$ delta function integral which sets $\omega' \to \omega\pm\Omega$. This gives
\begin{align}
	\chi_\pm^{(1)}(t,t') &= -\frac{1}{2}\delta f\omega_0^2 e^{\pm i\Omega t'}\int \frac{d\omega}{2\pi} e^{-i\omega(t-t')} \frac{1}{(\omega_0^2 - \omega^2 - i\omega\Gamma)}\frac{1}{(\omega_0^2 - (\omega\pm\Omega)^2 - i(\omega\pm\Omega)\Gamma)} \\
			&= -\frac{1}{2}\delta f\omega_0^2 e^{\pm i\Omega t'}\int d\tau\, e^{\pm i\Omega\tau} \chi^{(0)}(t-t'-\tau)\chi^{(0)}(\tau) \\
			&= -\frac{\delta f\omega_0^2 e^{\pm i \Omega t'} e^{-\frac{\Gamma}{2}(t-t')}}{2\omega_r^2}\int_0^\infty d\tau\, \theta(t-t'-\tau) e^{\pm i \Omega \tau}\sin[\omega_r(t-t'-\tau)]\sin[\omega_r\tau],
\end{align}	
where we have used $\omega_r = \frac{1}{2}\sqrt{4\omega_0^2-\Gamma^2}$, and taken advantage of convolutions, shifting frequency arguments, and the unperturbed harmonic oscillator response function $\chi^{(0)}(t-t')$. Performing the integration, and then adding the positive and negative frequency contribution as $\chi^{(1)}(t,t') = \chi_+^{(1)}(t,t') + \chi_-^{(1)}(t,t')$ gives the final result
\begin{align}
\chi^{(1)}(t,t') &= \frac{2 \delta f \omega_0^2}{\Omega  \omega_r \left(4\omega_r^2 - \Omega ^2\right)} \theta(t-t') e^{-\frac{1}{2} \Gamma  (t-t')} \cos \left(\frac{\Omega}{2}(t+t')\right) \\ &\times\left[2\omega_r \sin\left(\frac{\Omega}{2} (t-t')\right) \cos \left(\omega_r (t-t')\right)-\Omega  \cos \left(\frac{\Omega}{2}(t-t')\right) \sin \left(\omega_r (t-t')\right)\right].
\end{align}
The expression has several encouraging features. First, the expression is manifestly causal with the $\theta(t-t')$ dependence. Secondly, all terms have an explicit $(t-t')$ dependence, \emph{except for the $\cos(\Omega(t+t')/2)$ term}, which comes explicitly from the broken time translation invariance of the system. Additionally, the expression contains explicit exponential decay set by the loss parameter $\Gamma$, just as $\chi^{(0)}(t-t')$ does. 

Finally, we discuss the case of parametric resonance. The denominator clearly diverges when $\Omega = 2\omega_r$, which is the ubiquitous condition for parametric resonance that should be expected. When $\Omega \to 2\omega_r$, both the resonant denominator and the sum of terms in the numerator approach zero. Formally taking the limit recovers an important behavior
\begin{equation}
	\lim_{\Omega\to 2\omega_r} \chi^{(1)}(t,t') = \frac{\delta f \omega_0^2}{8\omega_r^3} \theta(t-t') e^{-\frac{\Gamma}{2}(t-t')} \cos(\omega_r(t+t'))\left(-2\omega_r(t-t') + \sin(2\omega_r(t-t'))\right).
\end{equation}
Most striking is the proportionality  $\propto \omega_r(t-t')$ which causes the function to grow until it eventual exponential suppression. It should be noted that around resonance, the actual time response of the system is sharp exponential growth, but this result captures the linear component of that response. 

\subsubsection{Quantum mechanical approach}

We now provide a quantum derivation of the Lorentz parametric oscillator susceptibility. We do this by calculating the dipole susceptibility $\alpha(t,t')$ for a quantum mechanical parametric oscillator, and showing that it matches the form of Eq. XX at first order in perturbation theory. We start with a Hamiltonian of a one dimensional quantum harmonic oscillator with an resonant frequency $\omega_0$ which oscillates at frequency $\Omega$.  
\begin{equation}
    H(t) = \frac{p^2}{2m} + \frac{1}{2}m\omega_0^2\left(1 + \delta\omega\cos\Omega t\right)x^2.
\end{equation}
In the basis of the unperturbed Harmonic oscillator, we can write this in terms of creation and annihilation operators as
\begin{equation}
    H(t) = \hbar\omega_0 a^\dagger a + \frac{\hbar\omega_0}{4}\delta\epsilon\cos\Omega t \left(a+a^\dagger\right)^2.
\end{equation}
The Floquet states, to first order in $\delta\epsilon$, are given as 
\begin{equation}
    \ket{\psi_n^{(1)}(t)} = e^{-i\omega_n t}\left[(1 + S_n)\ket{n} + P_{n-1}\ket{n-2} + - P_{n+1}^*\ket{n+2} \right]
\end{equation}
where we defined $S_n(t) \equiv -i\delta\epsilon (2n+1)\eta_0(t)$ and $P_n(t) \equiv -i\delta\epsilon\sqrt{n(n+1)}\eta_1(t)$, where
\begin{equation}
    \eta_0(t) = \frac{\omega_0 \sin\Omega t}{4\Omega}, \hspace{1cm} \eta_1(t) = \frac{\omega_0}{4}\frac{\Omega\sin\Omega t - 2i\omega_0 \cos\Omega t}{\Omega^2-4\omega_0^2}
\end{equation}
Using this, we can write the unitary time evolution order to first order as 
\begin{equation}
    U(t) = \sum_n e^{-i\omega_n t}\left[(1 + S_n)\ket{n}\bra{n} + P_{n-1}\ket{n-2}\bra{n} - P_{n+1}^*\ket{n+2}\bra{n}\right].
\end{equation}
Then we need the dipole moment $d(a+a^\dagger)$ in the interaction picture (i.e. transformed by $U(t)$). Doing this, and discarding terms greater than first order in $\delta\epsilon$, we find
\begin{equation}
    U^\dagger(t) (a+a^\dagger) U(t) = \sum_{n}\sum_{k=\pm 1, \pm 3} e^{-i(\omega_n-\omega_{n+k})t} A_k \ket{n+k}\bra{n}
\end{equation}
where 
\begin{align}
    A_1 &= \left(1 + S_{n+1}^*\right)\left[\sqrt{n+1}\left(1+S_n\right) - \sqrt{n+2} P_{n+1}^*\right] + P_n^* \left[\sqrt{n}\left(1+S_n\right) + \sqrt{n-1}P_{n-1}\right] \\
    A_{-1} &= \left(1 + S_{n-1}^*\right)\left[\sqrt{n}\left(1+S_n\right) + \sqrt{n-1} P_{n-1}\right] - P_n \left[\sqrt{n+1}\left(1+S_n\right) - \sqrt{n+2}P_{n+1}^*\right] \\
    A_3 &= -\sqrt{n+3}P_{n+1}^* + P_{n+2}^*\left[\sqrt{n+1}(1+S_n) - \sqrt{n+2}P_{n+1}^*\right] \\
    A_{-3} &= \sqrt{n-2}P_{n-1} - P_{n-2}\left[\sqrt{n}(1+S_n) + \sqrt{n-1}P_{n-1}\right]
\end{align}
Then we find that
\begin{equation}
    \Braket{0|d_I(t)d_I(t')|0} = d^2 e^{-i\omega_0(t-t')}\left(1 + S_0^*(t) + S_1(t) - \sqrt{2} P_1(t)\right)\left(1 + S_0(t') + S_1^*(t') - \sqrt{2} P_1^*(t')\right)
\end{equation}
Several of these terms are second order in $\delta\epsilon$ and can be discarded. Doing this, the commutator for linear response is
\begin{equation}
     \Braket{0|\left[d_I(t,)d_I(t')\right]|0} = \left(e^{-i\omega_0(t-t')} - \text{c.c}\right) - 2i\delta\epsilon d^2 e^{-i\omega_0(t-t')}\left[\eta_0(t) - \eta_0(t') - \left(\eta_1(t) - \eta_1^*(t')\right)\right] - \text{c.c}
\end{equation}
Then the polarizability to first order can be expressed as 
\begin{equation}
    \alpha(t,t') = \alpha^{(0)}(t-t') + \alpha^{(1)}(t,t') + \mathcal{O}(\delta\epsilon^2)
\end{equation}
where
\begin{equation}
    \alpha^{(0)}(t-t') = \frac{2d^2}{\hbar}\theta(t-t')\sin\left(\omega_0(t-t')\right)
\end{equation}
and
\begin{equation}
\begin{split}
    \alpha^{(1)}(t, t') &= -\frac{4\delta\epsilon d^2 \omega_0^2}{\hbar\Omega \left(\Omega^2 - 4\omega_0^2\right)}\theta(t-t')\cos\left(\frac{\Omega}{2}(t+t')\right) \\
    &\times \left[2\omega_0\sin\left(\frac{\Omega}{2}(t-t')\right) \cos\left(\omega_0(t-t')\right) - \Omega \cos\left(\frac{\Omega}{2}(t-t')\right) \sin\left(\Omega_0(t-t')\right)  \right] 
\end{split}
\end{equation}
Under the assumption that there is no dissipation, this is equivalent to the Lorentz parametric oscillator model that was derived semiclassically. 

\section{Electrodynamics with time-varying materials}

\subsection{Power transfer for a dipole}

For a point dipole, the total energy radiated can be written over all times can be written as $U = \int_0^\infty d\omega P(\omega)$, where
\begin{equation}
    P(\omega) = -\frac{\omega}{\pi}\Im \left[d(\omega)E^*(\omega)\right],
\end{equation}
where $d(\omega)$ is the dipole moment, and $E(\omega)$ is the applied field. Assuming the point dipole is described by the polarizability $\alpha(\omega, \omega')$, we can write the energy dissipated per frequency as
\begin{equation}
    P(\omega) = -\frac{\omega}{\pi} \Im \int_{-\infty}^\infty \frac{d\omega'}{2\pi}E_i^*(\omega)\alpha_{ij}(\omega, \omega')E_j(\omega').
\end{equation}
If the time variation of the dipole is periodic, then we can expand $\alpha(\omega, \omega') = \sum_k \alpha_k(\omega) \cdot 2\pi\delta(\omega - \omega' - k\Omega_0)$ to obtain
\begin{equation}
    P(\omega) = -\frac{\omega}{\pi} \Im \sum_k \alpha_k(\omega) E(\omega - k\Omega_0) E^*(\omega)
\end{equation}
where we have assumed that the polarizability is isotropic. We see that for $k=0$, we get $|E(\omega)|^2$ as usual so that $\Im \alpha_0(\omega)$ gives the dependence. If we take $E(t) = E_0 \cos(\omega_p t - \phi)$ then $E(\omega) = E_0 \pi e^{-i\phi}\left[\delta(\omega + \omega_p) + \delta(\omega - \omega_p)e^{2i\phi}\right]$ so that we get
\begin{equation}
    \frac{U}{T} = -\frac{E_0^2 \omega_p}{2}\left\{\Im \alpha_0(\omega_p) + \sum_{k=1}^\infty \Im\left[\alpha_k(\omega_p)e^{-2i\phi}\right]\delta_{2\omega_p = k\Omega_0}\right\}
\end{equation}

\subsection{Polarizability in a periodic system}

We can now think about how the dipole moment responds to a probe field. Since we have a time-periodic system, the polarizability takes the form 
\begin{equation}
    \alpha(\omega, \omega') = \sum_{k=-\infty}^\infty \alpha_k(\omega)2\pi\delta(\omega-\omega'-k\Omega)
\end{equation}
In other words, a probe frequency and a response frequency must differ by some multiple $k$ of the Floquet frequency $\Omega$. Then we note that if we apply a monochromatic electric field probe $E(t) = E_0 \cos(\omega_p t)$, we have in frequency space 
\begin{equation}
    E(\omega) = \pi E_0\left[\delta(\omega - \omega_p) + \delta(\omega + \omega_p)\right].
\end{equation}
Then the dipole response is given by
\begin{align}
    \Braket{\delta d(\omega)} &= \int \frac{d\omega'}{2\pi} \alpha(\omega, \omega') E(\omega') \\
    &= \sum_k \alpha_k(\omega) E(\omega + k\Omega)
\end{align}

\subsection{Connecting $\alpha$ to $\chi$}

\subsection{Maxwell's Equations in Time Dependent Materials}

In this section, we will formulate Maxwell's equations in media which are time-periodic and dispersive. Then by making the assumption of a spatially homogeneous medium, we will be able to cast Maxwell's equations into a Floquet eigenvalue problem for the wavevectors and quasifrequencies which can propagate. By solving this problem numerically, we can compute band structures. 

In the absence of external sources, we can then write down the Maxwell equation for the electric field as
\begin{equation}
	\curl\curl\Ev(\rv, \omega) - \frac{\omega^2}{c^2}\int \frac{d\omega'}{2\pi} \varepsilon(\rv, \omega,\omega') \Ev(\rv, \omega') = 0.
\end{equation}
If we write epsilon as time-independent background $\varepsilon_{\text{bg}}(\omega)$ plus a perturbation $\Delta\chi(\omega,\omega')$ which results from a time dependence, we can write $\varepsilon(\rv, \omega, \omega') = \varepsilon_{\text{bg}}(\rv, \omega)[2\pi\delta(\omega-\omega')] + \Delta\chi(\rv, \omega,\omega')$. Substituting this into the Maxwell equation, we obtain the form
\begin{equation}
    \curl\curl\Ev(\rv, \omega) - \frac{\omega^2}{c^2}\varepsilon_{\text{bg}}(\rv, \omega)\Ev(\rv, \omega)
    = \frac{\omega^2}{c^2}\int \frac{d\omega'}{2\pi} \Delta\chi(\rv, \omega,\omega') \Ev(\rv, \omega').
\end{equation}
If the driving is periodic, and the medium is assumed be uniform in space, then the response function can be cast into the form
\begin{equation}
	\Delta\chi(\omega,\omega') = \sum_k \Delta\chi_k(\omega)2\pi\delta(\omega-\omega'-k\Omega_0).
\end{equation}
Substituting this into the Maxwell equation gives
\begin{equation}
	\curl\curl\Ev(\omega) - \frac{\omega^2}{c^2}\varepsilon_{\text{bg}}(\omega)\Ev(\omega) = \frac{\omega^2}{c^2}\sum_k \Delta\chi_k(\omega)\Ev(\omega-k\Omega_0).
\end{equation}
Assuming a bulk medium which can be spatially decomposed into plane waves, and invoking the Bloch-Floquet requirement for the time-dependent portion of the mode functions, we can write
\begin{equation}
	\Ev(\rv, t) = e^{i \kv\cdot\rv}\sum_n u_{\Omega n} e^{-i(\Omega+n\Omega_0)t} \implies \Ev(\rv,\omega) = e^{i \kv\cdot\rv} 2\pi\sum_n u_{\Omega n}\delta(\omega-\Omega-n\Omega_0).
\end{equation}
For convenience, we will define $\Omega_n \equiv \Omega + n\Omega_0$. Substituting this into the Maxwell equation, integrating both sides in $\omega$ to isolate frequency components, and relabeling sums gives the final central equation result
\begin{equation}
	\left[k^2 - \frac{\Omega_n^2}{c^2}\varepsilon_{\text{bg}}(\Omega_n)\right]u_n = \frac{\Omega_n^2}{c^2}\sum_m \Delta\chi_m(\Omega_n)u_{n-m}.
\end{equation}
Rewriting this as an eigenvalue problem for $k$ in terms of $\Omega$, we have
\begin{equation}
	\frac{\Omega_n^2}{c^2}\varepsilon_{\text{bg}}(\Omega_n)u_n + \frac{\Omega_n^2}{c^2}\sum_m \Delta\chi_m(\Omega_n)u_{n-m} = k_\Omega^2 u_n,
	\label{eq:maxwell_floquet_eigenvalue}
\end{equation}
which casts the dispersion as a relatively simple central equation eigenvalue problem. Eq. \ref{eq:maxwell_floquet_eigenvalue} can be easily implemented as a matrix eigenvalue problem, yielding the wavevector $k_\Omega$ and Floquet mode amplitudes $u_{\Omega n}$ at each quasifrequency $\Omega \in [-\Omega_0/2, \Omega_0/2)$.

\subsection{Reflection and Transmission}

Once the Maxwell Floquet solutions have been obtained, as described in the previous section, they can be used to solve classical optics problems. In this section, we show the illustrative example of a reflection-transmission problem of a monochromatic field incident on a slab of material characterized by a time-dependent linear response function.

We write the Floquet modes as
\begin{equation}
	u_{b,\tilde{\Omega}}(t) = \sum_m u_{b,\tilde{\Omega}}^{(m)} e^{-i\Omega_0 m t},
\end{equation}
where $b$ is a band index, and $\sum_m$ is a sum over harmonics of the driving frequency $\Omega_0$. Then we can consider an interface between two materials (1) and (2), where (1) for now is air, and (2) is a material described with this kind of formalism, and we assume we have solved for the bands. Furthermore, we assume that the electric field is polarized in the plane of the interface (s-polarized). 

In this analysis, it will be helpful to express the incident frequency as $\omega_0 = \tilde{\omega}_0 + s\Omega_0$, where $\tilde{\omega}_0$ is confined to lie in the first Brillouin zone $-\frac{\Omega_0}{2} < \tilde{\omega}_0 < \frac{\Omega_0}{2}$. In this case, $s$ is easily interpreted as a number of harmonics by which the true incoming frequency is offset from the BZ frequency that is seen in the time-dependent medium. Then we can write the incident field as 
\begin{equation}
	E^{\text{inc}}(x,t) = e^{ik_0 x}e^{-i\omega_0 t},
\end{equation}
where $\omega_0$ is the frequency of the incident plane wave, and $k_0 = k(\omega_0) = \omega_0/c$ is the vaccuum dispersion. For now, we assume that the reflected field can have any frequency, so we can write very generally
\begin{equation}
	E^{(\text{r})}(x,t) = \sum_{\omega > 0} r(\omega) e^{-ik(\omega)x} e^{-i\omega t}.
\end{equation}
Finally, the transmitted field is
\begin{align}
	E^{(\text{t})}(x,t) &= \sum_b t_b e^{iq_b(\tilde{\omega}_0 x} e^{-i\tilde{\omega}_0 t} u_{b,\tilde{\omega}_0}(t) \\
		&= \sum_{b,m} t_b u_{b,\tilde{\omega}_0}^{(m)} e^{iq_b(\tilde{\omega}_0) x} e^{-i(\tilde{\omega}_0 + m\Omega_0) t}
\end{align}

The boundary conditions at the interface are given by $E^{\text{inc}}(0,t) + E^{\text{r}}(x,t) = E^{\text{t}}(0,t)$ and
$\partial_x E^{\text{inc}}(x,t)|_{x=0} + \partial_x E^{\text{r}}(x,t)|_{x=0} = \partial_x E^{\text{t}}(x,t)|_{x=0}$.

We can write a matrix equation down for the boundary conditions. Now we have boundary conditions at $x=\pm L/2$. The first two rows are for contituity of the field at $\pm L/2$ respectively. The last two rows are for continuity of the derivative. 
\begin{pmatrix}
-I*e^{ik_0(-L/2)} & \mathbf{u}_b e^{ik_b(\tilde{\omega}_0)(-L/2)} & \mathbf{u}_b e^{-ik_b(\tilde{\omega}_0)(-L/2)} & \mathbf{0} \\
\mathbf{0} & \mathbf{u}_b e^{ik_b(\tilde{\omega}_0)(L/2)} & \mathbf{u}_b e^{-ik_b(\tilde{\omega}_0)(L/2)} & -I*e^{ik_0(L/2)} \\
(\Omega_{m}/c)e^{i(\Omega_m/c)(-L/2)} & k_b(\tilde{\omega_0})\mathbf{u}_be^{ik_b(\tilde{\omega}_0)(-L/2)} & -k_b(\tilde{\omega_0})\mathbf{u}_be^{-ik_b(\tilde{\omega}_0)(-L/2)} & \mathbf{0} \\
\mathbf{0} & k_b(\tilde{\omega_0})\mathbf{u}_be^{ik_b(\tilde{\omega}_0)(L/2)} & -k_b(\tilde{\omega_0})\mathbf{u}_be^{-ik_b(\tilde{\omega}_0)(L/2)} & (\Omega_{m}/c)e^{i(\Omega_m/c)(L/2)}
\end{pmatrix}
\begin{pmatrix}
    \mathbf{r} \\ \mathbf{a} \\ \mathbf{b} \\ \mathbf{t}
\end{pmatrix} = 
\begin{pmatrix}
    \mathbf{v}_1 \\ \mathbf{v}_2 \\ \mathbf{v}_3 \\ \mathbf{v}_4
\end{pmatrix}

Finally, we have $\mathbf{v}_2 = \mathbf{v}_4 = 0$. Then we have $(\mathbf{v}_1)_s = e^{ik_0(-L/2)}$ and $(\mathbf{v}_3)_s = (\Omega_s/c)e^{i(\Omega_s/c)(-L/2)}$, where the $s$ index refers to the index that matches $s$ (so actually $M+1+s$).